\documentclass[a4paper,12pt,times,numbered,print,index,biblatex]{report}

\usepackage{thesis}
\usepackage{ifthen}
\usepackage{datetime}
\usepackage{amsmath,amssymb}
\usepackage[dvipsnames]{xcolor}

\usepackage[T1]{fontenc}
\usepackage{ae,aecompl}
\usepackage{graphicx}	
\usepackage{amsmath}
\usepackage{amssymb}
\usepackage[utf8]{inputenc} 
\usepackage[T1]{fontenc}    
\usepackage{hyperref}       
\usepackage{url}            
\usepackage{rotating}
\usepackage{indentfirst}
\usepackage{cases}
\usepackage{babel}
\newcommand{\sty}{\scriptstyle}
\newcommand{\ssty}{\scriptscriptstyle}
\newcommand{\tsty}{\textstyle}
\newcommand{\be}{\begin{equation}}
\newcommand{\ee}{\end{equation}}
\newcommand{\lb}[1]{\label{#1}}
\newcommand{\dl}{d_{\ssty L}}

\newcommand{\dg}{d_{\ssty G}}
\newcommand{\dz}{d_z}
\newcommand{\gaml}{\gamma_{\ssty L}}

\newcommand{\gamg}{\gamma_{\ssty G}}
\newcommand{\gamz}{\gamma_{\ssty z}}

\newcommand{\dobs}{d_{{\tsty {\ssty \rm obs}}}}
\newcommand{\Nobs}{N_{{\tsty {\ssty \rm obs}}}}
\newcommand{\Vobs}{V_{{\tsty {\ssty \rm obs}}}}
\newcommand{\gobs}{\gamma_{{\tsty {\ssty \rm obs}}}}

\newcommand{\doctitle}{} 
\newcommand{\doctype}{Galaxy Distribution Systems as Fractals}
% ...

\newcommand{\authorName}{Sharon Teles de Azevedo Chaves}

\newcommand{\supervisor}{Dr. Marcelo Byrro Ribeiro}

\newcommand{\cosupervisor}{Dr. Amanda Reis Lopes}

\newcommand{\Year}{\the\year{}}

\newcommand{\keywords}{Important, comma, separated, keywords, applicable, to, your, thesis}

\hypersetup
{
    pdfauthor={\authorName},
    pdfsubject={\doctitle},
    pdfkeywords={\keywords}
}

\begin{document}


\begin{titlepage}
\headheight = 0pt

{%
\fontsize{9pt}{9pt}\selectfont%
\renewcommand\arraystretch{2}

%
}

\vspace*{3.3 cm}

\begin{center}

{
\fontsize{33pt}{33pt}\selectfont%
\textsc{\textbf{\doctype}}
}

\vspace*{2.5 cm}
\begin{center}
Undergraduate dissertation submitted to the Valongo Observatory, Center for Natural and Mathematical Sciences, Universidade Federal do Rio de Janeiro, in partial fulfillment of the requirements for the degree of BSc in Astronomy\\   
\end{center}

\vspace*{2.5 cm}
\begin{Large}
\textbf{\authorName}
\end{Large}
\\
Orcid: 0000-0003-4497-9161
\vspace*{1.5 cm}

\begin{longtable*}{ p{0.5\textwidth} p{0.2\textwidth} }
\tabularnewline\supervisor & 1st Adviser
\ifthenelse{\equal{\cosupervisor}{Mr. Co-Supervisor's Name}}{}{%
\tabularnewline\cosupervisor & 2nd Adviser}
\tabularnewline
\end{longtable*}

\vspace*{1.5 cm}
13 September 2023
    
\end{center}
\end{titlepage}

\setcounter{page}{0}
\pagenumbering{arabic}   
\addtocounter{page}{1}
\normalsize

\chapter*{\centerline{Declaration of Originality}}\label{chapter:declaration}
I hereby certify that I am the sole author of this dissertation. All used materials, references
to the literature and the work of others have been referred to. 
The results presented here are original and were published in Refs. \cite{teles2020} and \cite{teles2022}.
This dissertation has not been
presented for examination anywhere else.

\pagebreak

\chapter*{\centerline{Dedication}}\label{chapter:dedication}
\begin{flushright}

\hspace*{29mm} Dedico esse trabalho às mulheres da história da minha família que não tiveram oportunidade de ser cientistas, e a todas que um dia serão.

\hspace*{30mm} I dedicate this work to the women in my family's history who did not have the opportunity to be scientists, and to all who will one day be.

\end{flushright}
\pagebreak

\chapter*{\centerline{Acknowledgments}}\label{chapter:acknowledgments}
To carry out this dissertation I relied on the help of several people. Among them, I would like to thank my first adviser, Dr. Marcelo Byrro Ribeiro, for correcting my work and teaching me throughout my university years. I am also grateful to my second adviser Dr. Amanda Reis Lopes for all the programming instructions that led me to discover and learn my favorite part of science. And to both of them, I am grateful for all the face-to-face and online meetings in which they were always willing to teach in a carefree way.
I am also grateful to the helpful professors and classmates with whom I collaborated during the course.
And I feel infinitely grateful to my friends Natanael and Paola, who helped me not to give up in moments of anxiety, and especially to my friend Erick who encouraged and supported me every day of the writing process.
I also thank my family's help, my sister, Sara Kimberlly, my mother, Eloisa, and my father, Aldinei, for believing in me since I was a kid and for always saying how smart and capable of anything I was.
Finally, I am grateful for the financial support from CNPq with my scientific initiation and to Universidade Federal do Rio de Janeiro for the scholarship which allowed me to finish my undergraduate studies.

\chapter*{\centerline{Agradecimentos}}\label{chapter:agradecimentos}
Para executar essa dissertação eu contei com a ajuda de diversas pessoas. Dentre elas, gostaria de agradecer ao meu orientador, Professor Doutor Marcelo Byrro Ribeiro, pelas correções ao meu trabalho e pelos ensinamentos ao longo dos anos de faculdade. Sou grata também à minha segunda orientadora, Doutora Amanda Reis Lopes, por todas as instruções em programação que me direcionaram a descobrir e aprender minha parte preferida da ciência. E aos dois sinto gratidão por todas as reuniões presenciais e online onde sempre tiveram disposição para ensinar de forma descontraída.
Também sou grata aos professores solícitos e colegas de classe com quem colaborei durante o curso.
E me sinto infinitamente agradecida aos meus amigos Natanael e Paola, que me ajudaram a não desistir nos momentos de ansiedade e em especial ao meu amigo Erick, que incentivou e apoiou todos os dias do processo de escrita.
Agradeço também à ajuda da minha família, minha irmã Sara Kimberlly, minha mãe Eloisa, e meu pai Aldinei, por acreditarem em mim desde quando eu era criança e sempre dizerem o quanto sou inteligente e capaz de qualquer coisa.
Por fim, sou grata ao suporte financeiro da CNPq com minha iniciação científica e à Universidade Federal do Rio de Janeiro pela bolsa de auxílio, sem os quais eu não teria conseguido terminar a faculdade.



\pagebreak

\phantomsection
\setcounter{tocdepth}{2}    
\renewcommand{\contentsname}{Table of Contents}
\tableofcontents

\clearpage \phantomsection
\setcounter{figure}{0}
\addcontentsline{toc}{chapter}{\listfigurename}
\listoffigures

\clearpage \phantomsection
\addcontentsline{toc}{chapter}{\listtablename}
\listoftables



\chapter*{\centerline{Abstract}}\label{chapter:abstract}
\addcontentsline{toc}{chapter}{Abstract}
This dissertation seeks to test if the large-scale galaxy distribution can be characterized as a fractal system. The standard $\Lambda$CDM cosmology having $H_0 = (70 \pm 5)$ km/s/Mpc is adopted on the study of almost a million objects from the UltraVISTA DR1, COSMOS2015 and SPLASH surveys, alongside the number density equations required for the description of these galaxy distribution systems as fractals with dimension D. The luminosity distance $d_L$, redshift distance $d_Z$ and galaxy area distance (transverse comoving distance) $d_G$ are the relativistic distance definitions used to estimate the galaxy number densities in the redshift interval of $0.1 \leq z \leq 4$ at volume limited subsamples defined by absolute magnitudes in the J-band for the UltraVISTA DR1 data and in the K-band for the COSMOS2015 and SPLASH data. Applying the appropriate relations for the description of galaxy fractal structures with single dimension $D$ in the relativistic settings to these surveys datasets it is possible to state that for $z<1$ the UltraVISTA DR1 galaxies presented an average of $D = (1.58 \pm 0.20)$, the COSMOS2015 galaxies produced $D = (1.39 \pm 0.19)$ and the SPLASH galaxies generated $D = (1.00 \pm 0.12)$. For $1 \leq z \leq 4$ the dimensions respectively decreased to $D = (0.59 \pm 0.28)$, $D = 0.54^{+0.27}_{-0.26}$ and $D = 0.83^{+0.36}_{-0.37}$. These results are robust under the Hubble constant uncertainty assumed here. Analysis of blue and red galaxies subsamples in the COSMOS2015 and SPLASH surveys show that the fractal dimensions of blue galaxies present essentially no alteration from the values above, although the ones for the red galaxies changed mostly to smaller values, meaning that D may be assumed as a more intrinsic property of the distribution of objects in the Universe, thus allowing for the fractal dimension to be used as a tool to study different populations of galaxies. All results confirm the decades old theoretical prediction of a decrease in the fractal dimension for $z>1$ suggesting that either there are yet unclear observational biases causing such decrease in the fractal dimension, or the galaxy clustering was possibly more sparse and the universe void dominated in a not too distant past.

\chapter*{\centerline{Resumo}}\label{chapter:resumo}
\addcontentsline{toc}{chapter}{Resumo}
Esta dissertação busca testar se a distribuição de galáxias em grande escala pode ser caracterizada como um sistema fractal. A cosmologia padrão $\Lambda$CDM com constante de Hubble $H_0 = (70 \pm 5)$ km/s/Mpc é adotada neste estudo de quase um milhão de objetos dos levantamentos UltraVISTA DR1, COSMOS2015 e SPLASH, juntamente com equações de densidade numérica necessárias para a descrição desses sistemas de distribuição de galáxias como fractais com dimensão $D$. A distância de luminosidade $d_L$, distância de desvio para o vermelho $d_Z$ e distância por área da galáxia (distância comóvel transversal) $d_G$ são usadas como definições de distância relativística para estimar as densidades numéricas de galáxias no intervalo de desvio para o vermelho de $0.1 \leq z \leq 4$ em subamostras limitadas por volume definidas por magnitudes absolutas na banda J para os dados do UltraVISTA DR1 e na banda K para os dados do COSMOS2015 e SPLASH. Aplicando aos conjuntos de dados desses levantamentos as relações apropriadas para a descrição de estruturas fractais de galáxias com dimensão única $D$ nas configurações relativísticas é possível afirmar que para $z<1$ as galáxias do UltraVISTA DR1 apresentaram em média $D = (1.58 \pm 0.20 )$, as galáxias do COSMOS2015 produziram $D = (1.39 \pm 0.19)$ e as galáxias do SPLASH geraram $D = (1.00 \pm 0.12)$. No intervalo $1 \leq z \leq 4$ as dimensões diminuíram respectivamente para $D = (0.59 \pm 0.28)$, $D = 0.54^{+0.27}_{-0.26}$ e $D = 0.83^{+ 0.36}_{-0.37}$. Esses resultados são robustos sob a incerteza na constante de Hubble assumida aqui. Análises de subamostras de galáxias azuis e vermelhas nos levantamentos COSMOS2015 e SPLASH surveys mostram que as dimensões fractais das galáxias azuis praticamente não apresentam alterações às mencionadas acima, embora as dimensões fractais das galáxias vermelhas tenham mudado principalmente para valores menores de desvio para o vermelho, o que significa que $D$ pode ser assumido como uma propriedade intrínseca da distribuição dos objetos no Universo, permitindo assim que a dimensão fractal seja utilizada como ferramenta para estudar diferentes populações de galáxias. Todos os resultados confirmam a previsão teórica de décadas de uma diminuição na dimensão fractal para $z>1$ sugerindo que ainda existem vieses observacionais causando tal diminuição na dimensão fractal, ou o agrupamento de galáxias era possivelmente mais esparso e o universo dominado por vazios em um passado não muito distante.

\chapter*{Introduction}\lb{intro}
\addcontentsline{toc}{chapter}{Introduction}
Several recent redshift galactic surveys display the large scale structure of the universe showing voids that shatter the luminous matter in an apparent irregular pattern, which combined with the cosmic microwave background subtle inhomogeneity can lead to ponder about some underlying assumptions of the standard $\Lambda$CDM cosmological model. Thus, it was brought to light the importance of defining the model scale limits and of testing the assumed homogeneity through observations to understand the evolution of the structure and get a glimpse of the universe's initial conditions. Analysis of the galaxies in these redshift surveys suggest that the distribution is irregular, possibly self similar and, therefore, fractal, which means that they can be represented by a power law. Self similarity is when the structure does not change with the scaling factor. Hence the distribution looks the same from afar or zoomed in. Additionally, the structure's irregularity can be characterized by the concept of fractal dimension which is a fractional value in the exponent of a power law. The mathematician who studied irregular objects with this approach was Benoît Mandelbrot, who from the 1950s to 1970s established adequate techniques to represent irregular shapes and forms by means of the fractal concept.

The fractal idea applied to galaxy distribution was originally known as the \textit{hierarchical cosmology} a century ago. The first attempt to mathematically describe the possible star hierarchy was published by E. Fournier d’Albe who proposed that the stars were infinitely distributed in a hierarchy of spheres in space where the mass inside a sphere was directly proportional to its radius $M(R) = R$ and the average matter density was zero \cite{albe1907}. In 1929, M. Amoroso Costa admitted the general relativistic implications on the universe spatial finitude. In his publication, Amoroso Costa investigated the possibilities of average density, potential and gravitational force for an one-dimensional star system both in a finite and infinite universe \cite{amoroso29}. Succeeding that there was a publication by Carpenter who detected that the distribution of the most luminous nebulae was not random but followed a specific regularity \cite{carpenter38}. However, the concept of hierarchy was lacking adequate mathematical representation and as a consequence Carpenter's results and the hierarchical cosmology were abandoned until 1970 when G. de Vaucouleurs and James Wertz reinterpreted them. Their suggestion pointed to an universal size-density power law existence that later was found to be in agreement with Mandelbrot's approach in his theory of fractals \cite{mandelbrot1977}. In 1980 the first redshift galaxy catalogs emerged and using the CfA redshift survey data de Lapparent et al. published evidence that the large scale structure of the universe did not appear as a smooth homogeneous distribution \cite{delapparent1986}. In the following years other authors confirmed the power law behaviour in the galaxy distribution to even deeper distances and some authors concluded that the universe could be described by a simple fractal, a succession of single fractal dimensions or even by a multifractal spectrum of dimensions \cite{ribeiro92a,pietronero87,jones88,balian88,martinez90}. 

Up to this point we discussed the cosmology only in the Newtonian context, however, a complete cosmology study needs general relativity. In this theory space and time are connected forming the space-time which is curved by the gravitational effects of the matter distribution. The object's trajectory in the curved space-time is described by a line called geodesic, thus, the light that was emitted by a galaxy in the past travels through the geodesic with constant speed $c$ forming the past light cone. There is also the future light cone formed by the light that objects are emitting now, however we can not observe it. Another important general relativistic aspect is that distance is not a unique concept because it depends on how it is measured. G.R.F. Ellis \cite{ellis71} observational relativistic distance definitions along with the ideas for density-size distribution considered by Wertz and Pietronero \cite{pietronero87,wertz70} were reviewed by Ribeiro \cite{ribeiro92a} who presented a relativistic model for the universe large scale structure as a self similar fractal system in a Lemaître-Tolman-Bondi (LTB) space-time. The analysis in Ribeiro \cite{ribeiro95} recognized that the spatial homogeneity of the Einstein-de Sitter cosmology (EdS) was observed only for nearby redshift values of $z \sim 0.01$, which suggests that the perceived fractal structure could be extended in the past light cone to higher $z$ values. In this way, the fractal distribution was not conflicting with the Cosmological Principle with regard to spatial homogeneity because general relativity allows us to distinguish between two types of homogeneity: observational homogeneity, represented in the past light cone where astronomical measurements are made, and spatial homogeneity, defined on the spatial hyper-surface where the density-energy term of Einstein's equation is constant in the standard cosmological model of Friedmann-Lemaître-Robertson-Walker (FLRW). Therefore, fractal galaxy distribution regards observational inhomogeneity.

Rangel Lemos \& Ribeiro \cite{juracy2008} analyzed the consequences of distinguishing spatial and observational homogeneity simulating the galaxy number counts behaviour taking into consideration the four cosmological distances: luminosity distance $d_L$, observer area distance $d_A$, galactic area distance $d_G$ and redshift distance $d_z$. Furthermore, they found the dependence between the radial densities and the distance definitions, and also the tendency of the fractal dimension to approach 3 as $z \rightarrow 0$, for all distances, and the asymptotic tendency of it going to zero as $z \rightarrow \infty$. This analyses was carried out using the spatially homogeneous EdS model and led to the curious result that spatially homogeneous cosmologies do not necessarily present observational homogeneity and, therefore, there is not contradiction with spatially homogeneous cosmological models like the FLRW which presents observational inhomogeneities at high redshift ranges \cite{juracy2008}. With these results the fractal dimension calculated from the observable quantities show that the relativistic fractal model can be a valid tool to characterize the galaxy distribution irregularity without conflicting with the standard model postulates. Additionally, obtaining more than one fractal dimension value in the same scaling range suggests that the distribution could have various values for the single fractal dimensions $D$ at different scaling ranges. 

Of notable interest to this dissertation is the study presented by Conde-Saavedra et al., who tested the fractal galaxy distribution hypothesis using the FORS Deep Field (FDF) dataset consisting of 5558 galaxies in the range $0.45 \leq z \leq 5.0$ \cite{gabriela}. They concluded that at $z \lesssim 1.3 - 1.9$ the sample produced an average single fractal dimension of $D = 1.4_{-0.6}^{+0.7}$ while beyond this threshold they obtained $D = 0.5_{-0.4}^{+1.2}$. Although the volume-limited samples are acquired from an indirect luminosity function method that produced high data uncertainties, this work contributed with further observational support for the theoretically predicted decreasing in fractal dimension at larger scales mentioned here. In this dissertation we followed this line of investigation by using volume-limited samples directly obtained from UltraVISTA DR1 \cite{ultra1}, COSMOS2015 \cite{La2016} and SPLASH \cite{Mehta2018} data, assuming a FLWR cosmological model. The determination of $D$ occurs by taking theses galaxy datasets with volume limited samples and plotting the number density vs. distance. Comparing with previous results, ours included better defined thresholds for moderate and high scaling ranges, smaller uncertainties and final results more in line with each other considering all cosmological distance definitions. 

Our calculations showed that in all three surveys the galaxy distribution can be well characterized as a fractal system possessing two consecutive scaling ranges. The UltraVISTA DR1 dataset containing 219,300 objects presented the following median fractal dimensions and uncertainties: $D = (1.58 \pm 0.20)$ for $z<1$, and $D = (0.59 \pm 0.28)$ for $1 \leq z \leq 4$ \cite{teles2020} which is also in agreement with the early theoretical prediction of a decrease in the fractal dimension at larger scales \cite{ribeiro95}. Extending the study, the COSMOS2015 dataset including 578,379 objects generated $D = 1.39 \pm 0.19$ for $z < 1$ and $D = 0.54_{-0.26}^{+0.27}$ for $1 \leq z \leq 4$ and the SPLASH dataset consisting of 390,362 objects presented $D = 1.00 \pm 0.12$ for $z < 1$ and $D = 0.83_{-0.37}^{+0.36}$ for $1 \leq z \leq 4$. These numbers turned out to be robust under the adopted Hubble constant uncertainty of $H_0 = (70 \pm 5)$ km/s/Mpc and provided further empirical support that the galaxy distribution can be characterized by two subsequent fractal scaling ranges with no detectable transition to homogeneity up to the redshift limits of these surveys \cite{teles2022}. 

Additional subsamples were generated by selecting blue, star forming galaxies and red, quiescent, ones using color-color diagrams or specific star formation rates. The fractal dimension of blue galaxies turned out either unaltered or only marginally altered as compared to the unselected samples. However, the red galaxies had their fractal dimensions becoming evidently smaller in most cases, apart from the red COSMOS2015 data whose D values increased noticeably. Such results indicate that single fractal dimensions may be used as descriptors of galaxy distributions as well as tools to trace galaxy types and/or their evolutionary stages at different redshift ranges. Supplementarily, all results collected here provide extra empirical confirmation of the theoretical decrease prediction in the fractal dimension at redshift values higher than the unity \cite{ribeiro95} and that the galaxy distribution can be described by two fractal scaling ranges without any transition to homogeneity up to the redshift limits of the three surveys.

The plan of the dissertation is as follows. Chapter 1 goes through the basic concepts and tools needed to understand relativistic fractal geometry. Chapter 2 specifies the observational details of the UltraVISTA DR1, COSMOS2015 and SPLASH redshift surveys used in this work. Chapter 3 features the results of the fractal analysis of the galaxy distribution surveys. Conclusion is used to discuss the results. Appendix A presents the python code used for the galaxy data processing. Appendix B exhibits mathematical details for the number-distance relation, essential to the discussions presented in Chapter 1.


\chapter{Basic Concepts}\label{chapter:first_chapter}



This chapter presents a discussion about fractals and their application to the relativistic domain. Section 1.1 will present fractals' fundamental concepts and Sec. 1.2 their application to galaxy distribution in the context of relativistic cosmology. 

\section{Fractals}
\label{sec:visao_geral}

Benoit Mandelbrot (1924-2010) developed the concept of fractals under the viewpoint that nature is too complex to be described by classically idealized geometrical shapes like lines, planes and geometrically smooth solids. 
The original concept of fractals had the aim of calculating the structure's irregularity by using a parameter called fractal dimension D that represents how rough the structure is. This is useful because in our world there are sea waves, mountains, coastlines, clouds; everything surrounding us is irregular if we look close enough.

\begin{figure}[ht]
\centering
\includegraphics[width=90mm]{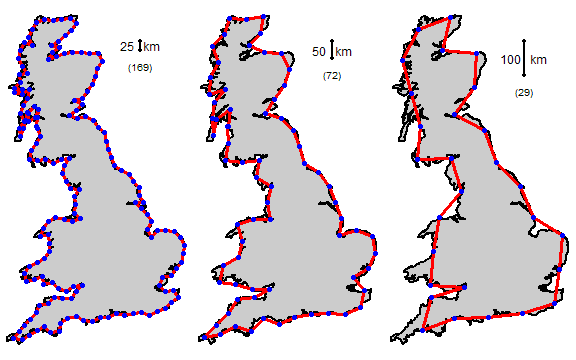}
    \caption[Different British coastline lengths]{Different British coastline lengths depending on the size of the measured reference line. On the left there are 169 straight lines 25 kilometers long. On the middle there are 72 straight lines 50 kilometers long. On the right there are 29 straight lines 100 kilometers long. The length of the British coastline would be 4,225 kilometers, 3,600 kilometers and 2,900 kilometers from left to right depending on the chosen yardstick. Figure from Ref. \cite{gurang2017}.}
    \label{fig:coastline11}
\end{figure}

Perhaps the first natural phenomenon studied by Mandelbrot was the length of coastlines, which is a useful illustration for our fractal analyses of the galaxy distribution \cite{mandelbrot67}. The idea is to be able to measure the length of a coastline using the concept of fractals, since it is not possible to find a unique number representation applying Euclidean geometry. This is because using our intuitive notion of length, a coastline would have many different extension values depending on how much detail it is being measured, and it would get to a point where if every ondulation is considered, the length of the coast would be infinite. In Fig. \ref{fig:coastline11} there is a drawing showing the different lengths the British coastline would have depending on how much detail is being measured using Euclidean geometry. On the left of Fig. \ref{fig:coastline11} there are 169 straight lines each being 25 kilometers long. The middle one has 72 straight lines 50 kilometers long, and the right one has 29 straight lines 100 kilometers long. From this we get the results that the length of the British coastline would be 4,225 kilometers, 3,600 kilometers and 2,900 kilometers from left to right in Fig. \ref{fig:coastline11}. This means we could decrease the size of the reference line as much as we wanted and only get to a relative value for the coastal length, since every time we look closer at an edge that is not smooth we will notice new curves. The fractal analyses is more practical due to the lack of this inconsistency problem as fractals have the important property of being self similar, meaning that they have the same irregularity level independent of measurement scales \cite{ribeiro98}. So, the British coastline would have an absolute value representation of its length, which would be proportional to its roughness.

\begin{figure}[ht]
\centering
	\includegraphics[width=80mm]{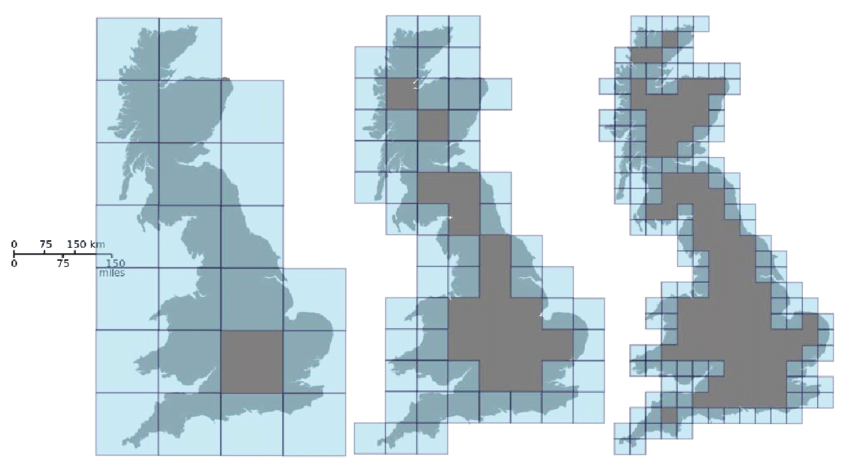}
    \caption[Box-counting method for the length of the British coastline.]{Box-counting method for the length of the British coastline. The boxes are scaled down from left to right. Figure from Ref. \cite{gurang2017}.}
    \label{fig:boxcounting}
\end{figure}

\begin{figure}[ht]
\centering
	\includegraphics[width=90mm]{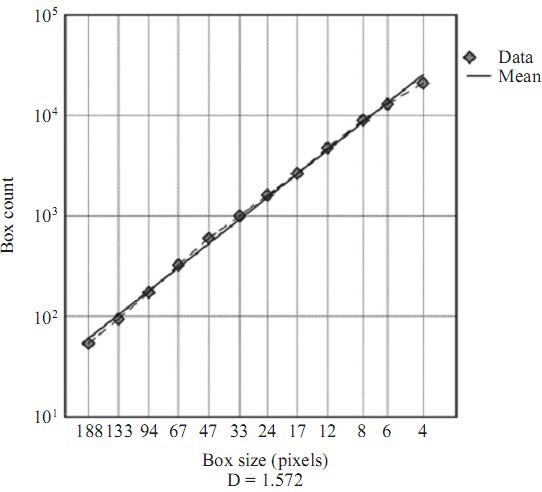}
    \caption[Log-log plot of the box-counting method for the length of the British coastline.]{Fractal dimension D of the British coastline from the log log plot of the number of boxes vs. the size of the boxes in the box-counting method. The fractal dimension of the British coastline is D = 1.572. Figure from Ref. \cite{gurang2017}.}
    \label{fig:britainlog}
\end{figure}

One of the methods that can be used to calculate the fractal dimension D is called box-counting. A grid is put over the map of Britain and the boxes touching the coastline are scaled up and down, as shown in Fig. \ref{fig:boxcounting}. From that, a plot is made using the number of boxes touching the coastline in the Y axis and the size of the box in the X axis generating the log log plot shown in Fig. \ref{fig:britainlog}. And finally, the slope of the adjusted line can lead directly to the fractal dimension D. It is important to mention that, as a comparison, if the box-counting method were to be applied to a smooth straight line the dimension found would be 1, and if applied to a smooth plane the dimension found would be 2.

\newpage
\section{Fractal Application to Galaxy Distributions}

\begin{figure}[H]
\centering
	\includegraphics[width=70mm]{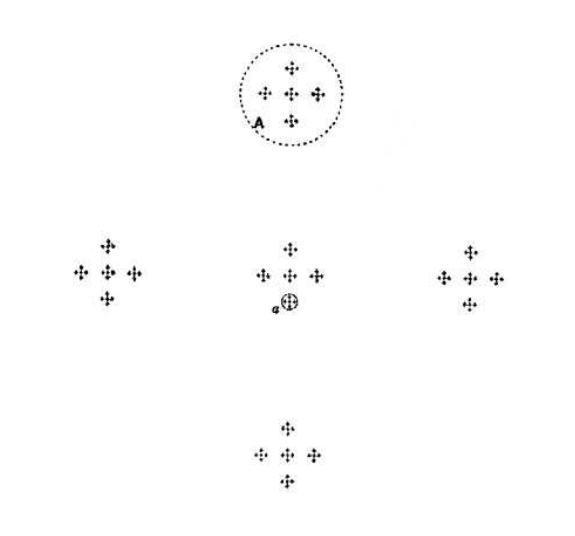}
    \caption[Star distribution by Fournier]{Star distribution by Fournier (1907) \cite{baryshev2002}.}
    \label{fig:fournier}
\end{figure}

The concept of what we modernly call fractal galaxy distribution actually started in the eighteenth century with Emanuel Swedenbord, who suggested in his book \textit{The Principia} that stellar systems form a non-uniform infinite hierarchical structure \cite{swedenborg1976principia}. In the book entitled \textit{Two New Worlds} \cite{albe1907}, Edmund Fournier d’Albe (1907) approached mathematically the description of stars as a hierarchy of spheres distributed in an infinite space, as shown in Fig. \ref{fig:fournier}, where the mass inside a given sphere is proportional to the radius of this sphere, $M(R) \propto R$, unlike the usual uniform approach where the mass is proportional do the radius cubed. 

Gérard de Vaucouleurs worked on a density-size law using multiple data and found that:

\begin{equation}
log \rho^*_M \simeq -1.9(log R - 10.7) ,     
\end{equation}

\noindent where $\rho^*_M$ is the average density of a galaxy cluster of radius R. In view of this, he advanced the following proposition without the logarithm \cite{vaucouleurs60}: 
\begin{equation}
M(R) \propto R^{3-D} , 
\end{equation}
at which the fractal dimension D is associated with the galaxy distribution's inhomogeneity degree.

At this point, the hierarchical concept was lacking mathematical formalism until the books \textit{Fractals: form, chance and dimension} and \textit{The fractal geometry of nature} \cite{mandelbrot1977,mandelbrot1982}, where Benoit Mandelbrot presented mathematical fractal properties for galaxy distribution, generalizing Einstein's cosmological principle, where D = 3 (spherically homogeneous distribution), to obtain a non-uniform distribution of galaxies with D < 3. Mandelbrot also proposed the “Conditional  Cosmological Principle” stating that observers would only perceive the same structure around them if they are observing from a material point like a galaxy but not in vacuum, so the detection of decrease in density would have to be scaled larger than the size of the vacuum \cite{mandelbrot2002}.

\begin{figure}[H]
\centering
	\includegraphics[width=100mm]{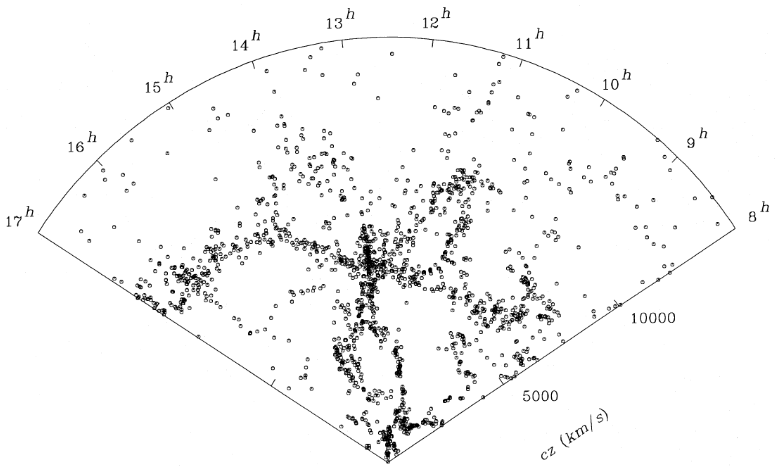}
    \caption[CfA Redshift Survey]{CfA Redshift Survey: Map of galaxy distribution containing 1761 galaxies \cite{cfa88}.}
    \label{fig:cfaredshift}
\end{figure}

Nevertheless, the galaxy distribution inhomogeneity had its most important observational support with the release of the CfA Redshift Survey in 1988 by means of a famous picture of the galaxy distribution as shown in Fig. \ref{fig:cfaredshift}.

Almost simultaneously, Luciano Pietronero proposed the conditional density $\Gamma$ as a scaled mean density since in a fractal system the density is dependent on position and volume observed \cite{pietronero87}. Such density was defined as

\begin{equation}
\Gamma(r) = \frac{D}{4 \pi} B r^{D-3} , 
\end{equation}

\noindent
where B is a scale parameter and r is the distance from the observer. 

\begin{figure}[H]
\centering
	\includegraphics[width=70mm]{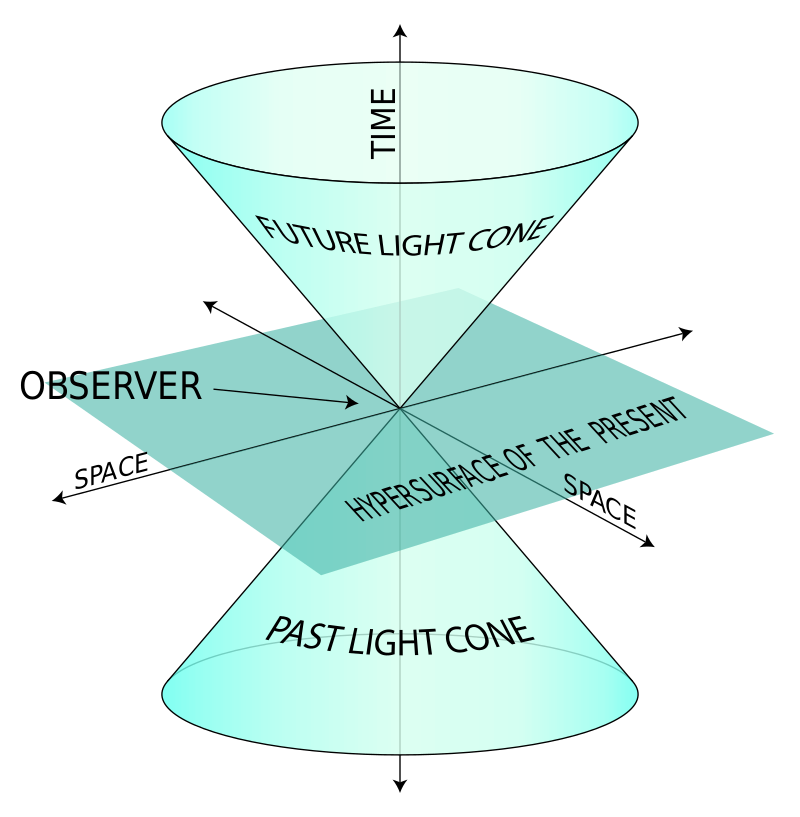}
    \caption[Relativistic light cone of observer]{Observer's light cone. The observer is in the origin of the cones as the lower one is the past light cone, which is the light observed from the past, while the upper one is the future light cone, representing the light that is emanating from the observer.}
    \label{fig:lightcone}
\end{figure}

Relativistic fractal cosmology considers the fact that the expanding universe is curved by the presence of matter, and so the distances are not absolute as in the Newtonian approach. Additionally, our universe can be represented by a light cone, which is the path a light beam would take through spacetime when moving in all directions, as shown in Fig. \ref{fig:lightcone}. 
The FLRW cosmology describes a homogeneous and isotropic universe, which is not in disagreement with a fractal model since the homogeneity of the FLRW cosmological model is spatial and the inhomogeneity of the fractal model is observational, that is, along the spacetime lightcone.

\subsection{Newtonian Cosmology}

In the Newtonian hierarchical fractal cosmology, $V_{obs}$ is defined as the observational volume, $\dobs$ is an observational distance, $\gobs^\ast$ is the observed volume number density and $\Nobs$ is the observed cumulative number counts of galaxies. From these terms we can write the following expressions,
\be
V_{obs} = \frac{4}{3}\pi (d_{obs})^3,
\lb{vobs}
\ee

\be
\gobs^\ast=\frac{\Nobs}{\Vobs}.
\lb{gobs-ast}
\ee

The \textit{Pietronero-Wertz hierarchical (fractal) cosmology} \cite{ribeiro98,ribeiro94} starts by assuming a critical phenomenological expression for the galaxy observed cumulative number counts $\Nobs$ that depends on the distance $\dobs$ \footnote{Since this is a galaxy distribution scale, the minimum observational distance is the order of the local galaxy group and the maximum distance will depend on the observational samples.} via the power law called the \textit{number-distance relation} given by

\be
\Nobs=B \, {(\dobs)}^{D}, 
\lb{Nobs}
\ee

\noindent where $B$ is a scale parameter and $D$ is the single fractal dimension.\footnote{See the Appendix B for the heuristic derivation of Eq. (\ref{Nobs})} Such relation gives the number of sources per unit of observational volume out to a distance $\dobs$. Additionally, $\Nobs$ is a radial quantity, so, it is not understood in statistical sense considering that it does not average all points against all points. From Eqs.\ (\ref{vobs}), (\ref{gobs-ast}) and (\ref{Nobs}) we define the \textit{De Vaucouleurs density power-law} $\gobs^\ast$ for a fractal structure as
\be
\gobs^\ast = \frac{\Nobs}{\Vobs}= \frac{3B}{4\pi}{(\dobs)}^{D-3},
\lb{gstar3}
\ee
where the term $3B / 4\pi$ is a positive constant \cite{pietronero87,ribeiro94}.

If in the previous expression $D=3$, $\Nobs$ increases with $\Vobs$ and galaxies would be evenly distributed along all regions of the observed space. However, if $D<3$, as $\dobs$ increases $\Nobs$ grows at a smaller rate than $\Vobs$, forming then gaps in the galactic distribution, in other words, regions devoid of galaxies. Close to these galactic gaps there would be regions where galaxies clump. Hence, voids and galactic clumpiness would be a consequence of a fractal galaxy structure whose fractal dimension is smaller than the topological dimension in which the galaxy structure is embedded. In this case the fractal dimension becomes a descriptor of galactic clumpiness or of void dominance in the galactic structure. 

Since $\Nobs$ is a cumulative quantity, it does not increase with $\dobs$ if beyond a certain distance there are no longer galaxies. However, if objects are still detected and counted, even at irregular rate, then $\Nobs$ continues to increase. This amount of growth can be influenced by observational biases, possibly leading to an irregular behavior, however $\Nobs$ is required to grow or remain constant and, so, the exponent in Eq.\ (\ref{Nobs}) must be positive or zero.
Therefore, if the galaxy distribution fractal dimension is $D<3$, then the observational number density decays as a power-law. If $D=3$ galaxies are uniformly distributed and the galactic structure is called \textit{observationally homogeneous}. In this instance the number density is constant and distance independent. Lower values of $D$ imply a more abrupt power-law decay and, as a consequence, more gaps in the galaxy distribution. 

\subsection{Relativistic Cosmology}

In relativistic cosmology the observer's past light cone is the geometrical locus of observations, which means that even spatially homogeneous cosmologies like the FLRW do not generate observationally constant number densities at moderate and high redshift ranges, since observational and spatial homogeneities are different relativistic concepts in cosmology. So, $\gobs^\ast$ is an average relativistic density. In addition, in general relativity distances may vary, especially for $z>1$, a situation which indicates that at high redshift $\dobs$ would have different values for each distance definition at the same redshift $z$. This means that as distance in cosmology is not uniquely defined $\dobs$ must be replaced by $d_i$ in the previous equations of this chapter \cite{ellis71,ellis2007,holanda2010,holanda2011,holanda2012}. 

The distance definitions applicable to relativistic fractal cosmologies are the luminosity distance $d_L$, the redshift distance $d_Z$ and the galaxy area distance $d_G$, also known as transverse comoving distance. The luminosity distance is defined according to, 

\begin{equation}
    F = \frac{L}{4 \pi d_{L}^{2}} \, ,
\end{equation}

\noindent where $F$ is the flux density of the object and $L$ is its total amount of electromagnetic energy emitted per unit of time. From an operational and numerical viewpoint, $d_L$ comes from the python parameter $\Lambda$CDM from the package astropy.cosmology with the following settings: Hubble constant $H_0 = 70 km \cdot s^{-1} \cdot Mpc^{-1}$, Matter density parameter $\Omega_{m0} = 0.3$ and Dark energy density parameter $\Omega_{\Lambda0} = 0.7$, for the FLRW cosmology assumed.

Using $d_L$ we defined $d_G$ considering the Etherington reciprocity law,
\be
\dg = \frac{\dl}{1+z}.
\lb{eth}
\ee
The redshift distance is defined as follows, 
\be 
\dz=\frac{c \, z}{H_0},
\lb{red}
\ee

\noindent
where $c$ is the speed of light in vacuum and $H_0$ is the Hubble constant. 
The previous expressions must be rewritten as below in order to become suitable for relativistic cosmologies, 
\be
\dobs=d_i,
\lb{dists}
\ee
\be
\Vobs=V_i=\frac{4}{3} \pi {(d_i)}^3,
\lb{vi}
\ee
\be
\Nobs=N_i=B_i \, {(d_i)}^{D_i}, 
\lb{Nobs_i}
\ee
\be
\gobs^\ast=\gamma^\ast_i =\frac{N_i}{V_i}=\frac{3B_i}{4\pi}
{(d_i)}^{D_i-3},
\lb{gobs-ast_i}
\ee

\noindent
where $i=({\ssty L}$, ${\sty z}$, ${\ssty G})$ corresponding to the chosen distance definition, which is attached to each specific constant $B_i$ and fractal dimension $D_i$. This is because $N_i$ is counted considering the limits given by every particular distance definition, so for a given $z$ each $d_i$ will generate its respective $V_i$, $N_i$,
$B_i$ and $D_i$. Therefore, every quantity is connected to a certain distance definition.

\noindent Subsequently, to calculate the fractal dimension $D_i$ it was necessary the linearization of the graph $\gamma^*_{i}$ vs. $d_i$ so we plotted the log-log results for $\gamma^*_{i}$ vs. $d_i$ to obtain a value for the slope of the points and used the relation in the following equation:
\begin{equation}
D_i = 3 + \;\mbox{slope}\; ,
\label{slope3}
\end{equation}
where $D_i$ \textit{(i = {\scriptsize L,Z,G})}.

\chapter{Galaxy Surveys Description}\label{chapter:second_chapter}
\begin{figure}[H]
\centering	
        \includegraphics[width=150mm]{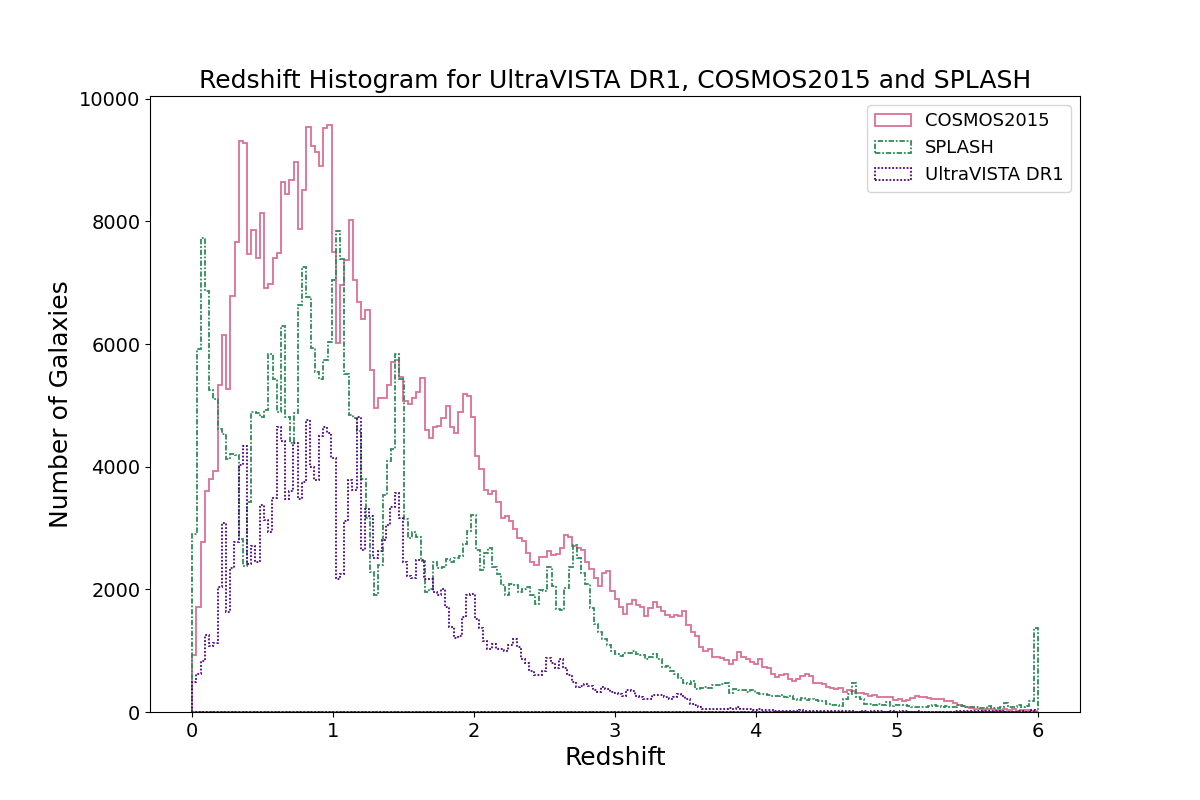}
    \caption[Histogram of galaxy distribution $vs$ redshift]{Histogram showing the galaxy distribution numbers in terms of redshift for the UltraVISTA DR1, COSMOS2015 and SPLASH surveys datasets studied here. This graph has $\Delta z \approx 0.03$ as the redshift bins’ size.}
    \label{fig:estatisticacatalogos}
\end{figure}

The three catalogues chosen for the study are presented in this chapter. There are two important features that were taken into account for their choice: high values of redshift, which is needed for our relativistic analysis, and a large number of galaxies, due to the statistical nature of the fractal analysis. In Fig. \ref{fig:estatisticacatalogos} there is a histogram showing the galaxy distribution numbers in terms of redshift for the UltraVISTA DR1 \cite{ultra1}, the COSMOS2015 \cite{La2016} and the SPLASH \cite{Mehta2018} surveys.

\section{UltraVISTA DR1}

\begin{figure}[H]
\centering
	\includegraphics[width=150mm]{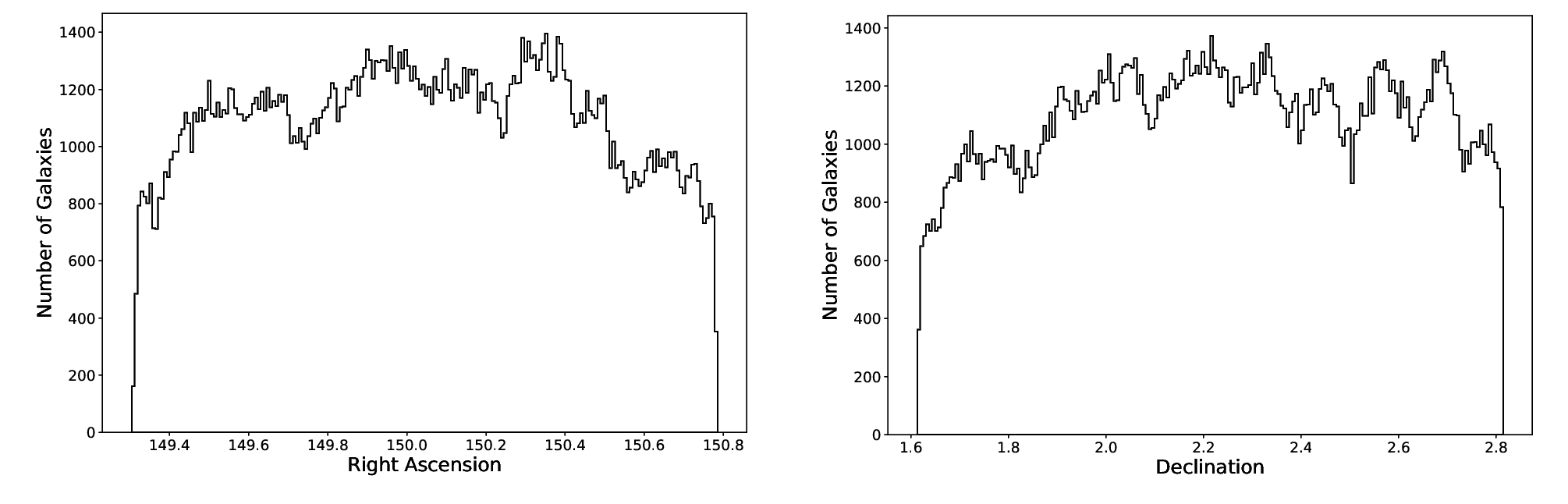}
    \caption[Galaxy distribution $vs$ right ascension and declination in UltraVISTA DR1]{On the left is the histogram showing the galaxy distribution numbers in terms of right ascension (deg) and on the right is the histogram showing the galaxy distribution numbers in terms of declination (deg), both for the UltraVISTA DR1 survey dataset studied here.}
    \label{fig:uvistadr1ceu}
\end{figure}

The first data release of the UltraVISTA galaxy survey is centered on the COSMOS field \cite{cosmos2007} with an effective area of $1.5\,\mbox{deg}^{\,2}$. The observations were made in four near infrared filters, $Y$, $J$, $H$ and $K_{\mathrm{S}}$, described in Ref. \cite{ultra1}.
After applying the SED fitting technique to 29 bands, the photometric redshifts were calculated by Ref. \cite{ultra11} using the ones from UltraVISTA and another set of broad and narrow bands from other surveys covering the mid infrared, near infrared, optical and ultraviolet regimes. 
The original dataset consisted in a sample of 219300 galaxies in the redshift range of $0.1\le z\le6$.  
In Fig. \ref{fig:uvistadr1ceu} there is a histogram of the galaxy distribution numbers in terms of right ascension and declination for the UltraVISTA DR1 survey, which are $1.61\le\mbox{DEC (deg)}\le 2.81$ and $149.30\le \mbox{RA (deg)}\le150.78$, with limiting apparent magnitude of $K_S=24.0$.

\section{COSMOS2015}

\begin{figure}[H]
\centering
	\includegraphics[width=150mm]{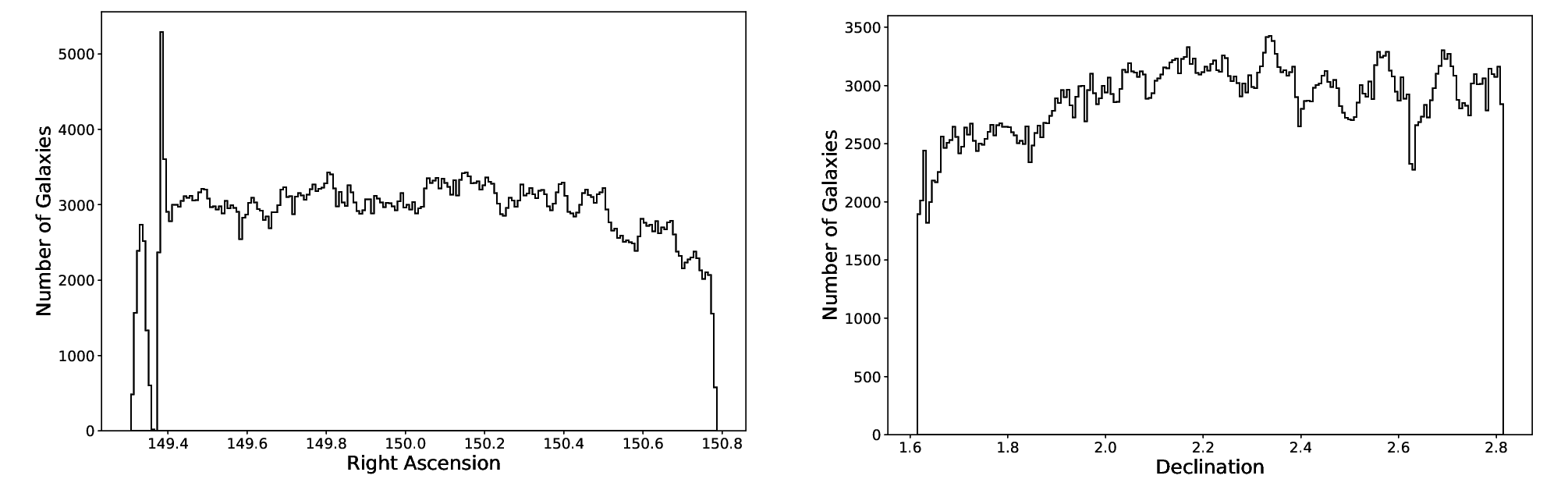}
    \caption[Galaxy distribution $vs$ right ascension and declination in COSMOS2015]{On the left is the histogram showing the galaxy distribution numbers in terms of right ascension (deg) and on the right is the histogram showing the galaxy distribution numbers in terms of declination (deg), both for the COSMOS2015 survey dataset studied here.}
    \label{fig:cosmos2015ceu}
\end{figure}

The COSMOS2015 catalog includes infrared data from the Spitzer Large Area survey over 2 deg$^2$ in the COSMOS field, Y-band images from Subaru/Hyper-Suprime-Cam and YJHK$_{S}$ images from the UltraVISTA DR2 \cite{cosmos2007}.
The UltraVISTA DR2 region contains 578379 galaxies where the object detection is performed by the $\chi^2$ sum of the YJHK$_{S}$ and z$^{++}$ images. This means that the sample has 359079 new objects as compared to the UltraVISTA DR1.
The observed area has the range of coordinates as ilustrated in Fig. \ref{fig:cosmos2015ceu}, which are: $1.43\le\mbox{DEC (deg)}\le 3.05$ and $149.28\le \mbox{RA (deg)}\le151.05$, with hundreds of thousands objects at redshift $0.1\le z\le6$ and limiting apparent magnitude of $K_S=24.0$ at $3\sigma$ in a 3" diameter aperture and parts of the field covered by the ``ultra-deep stripes'' (0.62 deg$^2$) have limiting apparent magnitude $K_S=24.7$ at $3\sigma$ in a 3" diameter. The photometric redshifts were computed using LePHARE 
\cite{arnouts,refId0} following Ref. \cite{ultra11}.

\section{SPLASH}

\begin{figure}[H]
\centering
	\includegraphics[width=150mm]{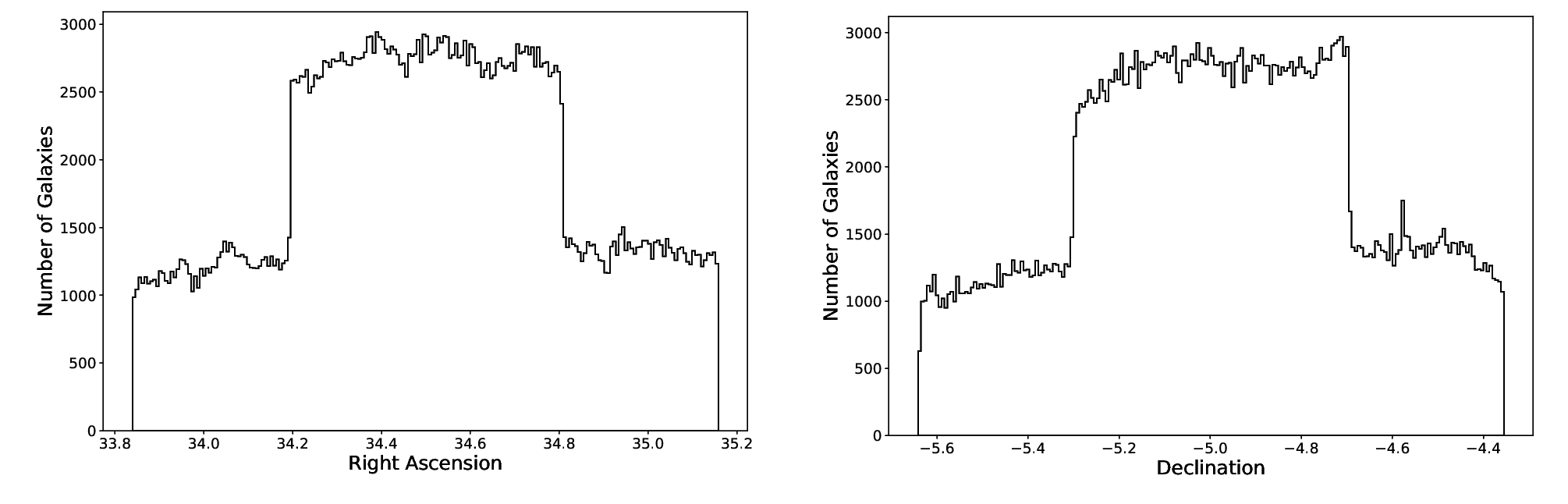}
    \caption[Galaxy distribution $vs$ right ascension and declination in SPLASH]{On the left is the histogram showing the galaxy distribution numbers in terms of right ascension (deg) and on the right is the histogram showing the galaxy distribution numbers in terms of declination (deg), both for the SPLASH survey dataset studied here.}
    \label{fig:splashceu}
\end{figure}

The SPLASH survey is a multi-wavelenght photometry catalog with coverage area of $2.4\,\mbox{deg}^{\,2}$. The sources were identified using a detection image defined as a $\chi^2$ combination of ugri images from CFHT Legacy Survey (CFHTLS), u image from Megacam Ultra-deep Survey: U-Band Imaging (MUSUBI), YJHKs images from VISTA Deep Extragalactic Observations (VIDEO), JHK images from Ultra Deep Survey (UDS) DR11 and grzy images from Hyper-Suprime-Cam (HSC) DR1. This survey follows a different strategy for the detection of the objects, as it includes the u-band, in order to recover the bluest objects.
This catalog contains 390362 galaxies at $0<z<6$, where the redshifts were measured using LePhare with a similar method to that used for the COSMOS field \cite{ultra11,La2016}.
In Fig. \ref{fig:splashceu} there is a histogram of the galaxy distribution numbers in terms of right ascension and declination for the SPLASH survey, which are $-6.04\le\mbox{DEC (deg)}\le-3.96$ and $33.45\le \mbox{RA (deg)}\le35.55$, with limiting apparent magnitude of $K=24.7$. 
The SPLASH galaxies were mapped in a different region of the sky and it follows a different strategy for object detection as compared to the other two surveys, which validates its inclusion in the fractal analyses, along with the sizable portion of galaxies in the redshift interval $1<z<4$ as shown in Fig. \ref{fig:estatisticacatalogos}.

\chapter{Fractal Analysis}\label{chapter:third_chapter}
As previously discussed, this study aims at testing whether or not the fractal galaxy distribution hypothesis holds at very large scales of the observed Universe, hence we chose to perform a fractal analysis with the data provided by the UltraVISTA DR1, COSMOS2015 and SPLASH galaxy surveys exactly because each of them contains hundreds of thousands of galaxies with measured high redshift values.

\section{UltraVISTA DR1}

\subsection{Data selection}
The establishment of volume limited samples is needed for the fractal analytical approach detailed in the first chapter. Knowing that galaxy surveys are limited by apparent magnitude, we tested if reduced subsamples would follow increasing redshift bins for the sake of rendering the reduced galaxy data distributed over limited volume bins. 
The data reduction strategy was to select galaxies below a certain absolute magnitude threshold determined by the limiting apparent magnitude on the survey. This was made on the plot of absolute magnitude in terms of their measured redshifts using the following expression: 
\be M=m-5\log d_L(z) -25, \lb{magabs1} \ee 
where $M$ is the absolute magnitude, $m$ is the apparent magnitude and $d_L$ is the luminosity distance given in Mpc. Afterwards we chose the apparent magnitude threshold $m$, its passband and verified if the resulting data is distributed along the measured redshift bins to establish a redshift window range for the final subsample distribution. 

\begin{figure}[H]
\centering
	\includegraphics[width=100mm]{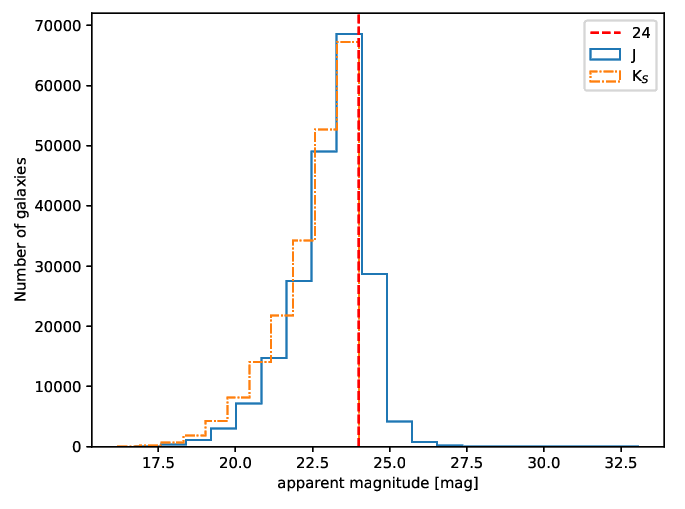}
    \caption[Histogram of galaxy distribution $vs$ apparent magnitude in UltraVISTA DR1]{Histogram of galaxy distribution $vs$ apparent magnitude (mag) for the UltraVISTA DR1 survey dataset studied here.}
    \label{fig:uvistadr1mag}
\end{figure}

\begin{figure}[H]
\centering
	\includegraphics[width=100mm]{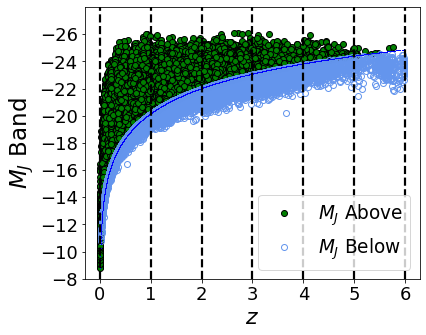}
    \caption[UltraVISTA DR1 absolute magnitude selection]{Plot of the absolute magnitudes for the UltraVISTA DR1 galaxies in terms of their photometrically measured redshift values. The dividing line corresponds to apparent magnitude $J = 24$ and all galaxies having $M_J$ above this cutoff and $z \leq 4$ were included in the subsample.}
    \label{fig:uvistaselection}
\end{figure}

The UltraVISTA DR1 survey provided absolute magnitudes calculated in the $NUV$, $B$, $R$ and $J$ passbands. Since the near infrared filter $J$ is less affected by dust extinction it is the best choice for our objectives here. In Fig.\ \ref{fig:uvistadr1mag} it is shown an histogram of the UltraVISTA DR1 galaxy numbers in terms of apparent magnitudes in the $K_{\mathrm{S}}$ and $J$ passbands where we can see that the number distribution peaks at the apparent magnitude value 24 for both wavebands, an information that led us to adopt $J=24$ as our apparent magnitude threshold. So, Eq.\ (\ref{magabs1}) can be rewritten according to the following expression, which provides the dividing line between the selected and unselected data.
\be M_J=24-5\log d_L(z) -25. \lb{magabs2} \ee
Fig.\ \ref{fig:uvistaselection} displays a plot of the 219300 UltraVISTA galaxies' absolute magnitudes in the $J$-band against their respective measured redshifts. Green filled circles are those whose absolute magnitudes supplied by the survey are smaller than $M_J$ in Eq.\ (\ref{magabs2}), and are then a part of the reduced subsample, while the open blue circles are outside this threshold.

Since the data generally follow the redshift bins of increasing values we noticed that the absolute magnitude cut based on the $J$-band is consistent with a volume-limited subsample. Additionally, we noticed that there are fewer galaxies at the tail of the distribution, so our subsample went through another cut at $z=4$. Hence, we came to a \textit{reduced UltraVISTA DR1 subsample} of 166566 galaxies cut by $J$-band absolute magnitudes $M_J$ and restricted up to $z=4$. The remaining 52734 galaxies outside this subsample, that is, below the dividing line in Fig.\ \ref{fig:uvistaselection} or having $z>4$, were not considered for further analysis.

\subsection{Data Analysis}

\begin{figure}[H]
\centering
	\includegraphics[width=170mm]{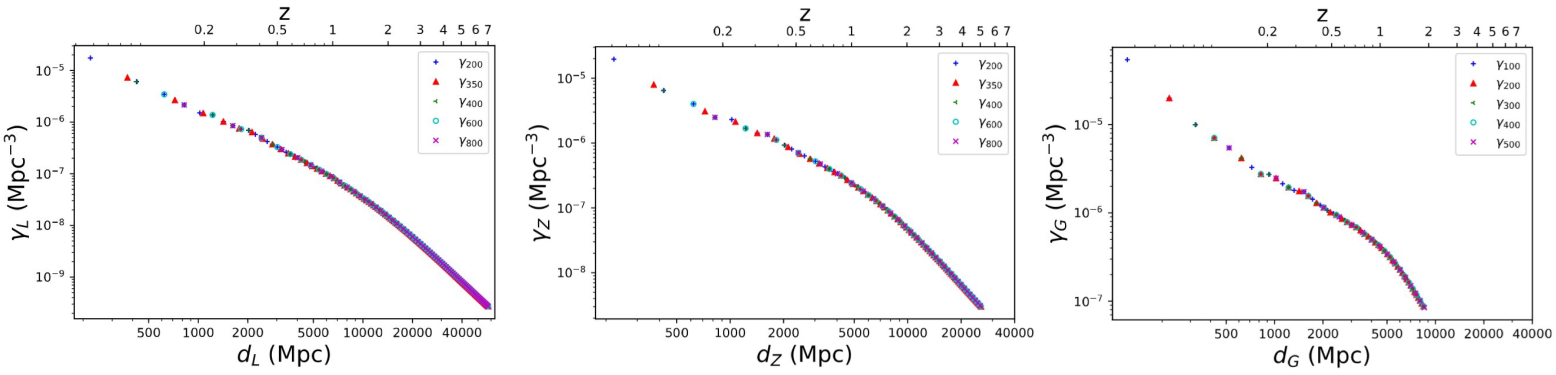}
    \caption[Plot of $\gamma^*_{i}$ vs. $d_i$ in the UltraVISTA DR1 sample for different distance bin values]{Graph showing the results for $\gamma^*_{L}$ vs. $d_L$, $\gamma^*_{Z}$ vs. $d_Z$ and $\gamma^*_{G}$ vs. $d_G$, from left to right, obtained with the UltraVISTA DR1 galaxy redshift survey dataset for distance bin values varying from 200Mpc to 800Mpc.}
    \label{fig:bins}
\end{figure}

In the python algorithm we chose to separate the data using a fixed distance interval $\Delta d_i = 200$Mpc, and from that, the volume $\Delta V_i$ was determined and the number of galaxies $N_i$ in this volume was calculated. We also defined the volume density $\gamma ^{*}_i$ as in Eq. (\ref{gobs-ast_i}). The algorithm in the Appendix A was designed to calculate the cumulative value of $\Delta d_i$, $\Delta V_i$, $N_i$ and $\gamma ^{*}_i$, so the first distance interval is up to 200Mpc away from earth, the second interval is up to 400Mpc and so on. 
We tested different fixed distance intervals in the data analyses to check if it would alter the distribution of the points in the graph $\gamma^*_{i}$ vs. $d_i$ and it did not, as seen in Fig. \ref{fig:bins}.

Now one needs to choose a cosmological model to acquire the observational distances $d_i\,(i={\ssty G}$,
${\ssty L}$, ${\sty z})$ from the calculated photometric redshift values. We assumed the FLRW cosmology with $\Omega_{m_{\ssty 0}}=0.3$, $\Omega_{\Lambda_{\ssty 0}} = 0.7$ and $H_0=70 \; \mbox{km} \; {\mbox{s}}^{-1} \; {\mbox{Mpc}}^{-1}$. 
The next steps were to establish the minimum redshift value $z_{\ssty 0}$ from where the analysis starts, the respective minimum distances $d_{i_0}=d_{i_0}(z_{\ssty 0})$, and the incremental distance interval $\Delta d_i$. 
The process of sorting out the data was initiated by counting the number of observed galaxies $N_{i_{\ssty \rm 1}}$ in the first interval between $d_{i_0}$ and $d_{i_1}=d_{i_0}+\Delta d_i$ and calculating the respective volume density $\gamma_{i_1}^\ast$. The first bin was defined by this first interval. Next, the size of the bin was increased by $\Delta d_i$ and cumulative values for $N_{i_{\ssty \rm 2}}$ and $\gamma_{i_2}^\ast$ were acquired at the distance $d_{i_2}=d_{i_0}+2\Delta d_i$. This algorithm was reproduced $n$ times until the farthest group of galaxies were included and all related quantities were also counted and calculated.
In order to find out whether or not different bin size increments $\Delta d_i$ affect the results we tested many values for each distance definition. This test turned out negative, meaning that the obtained results are independent of bin size increment. Then, we chose the value $\Delta d_i=200$~Mpc which was implemented in all calculations and presented a very reasonable amount of data points for each quantity from where simple linear regression analyses were ready to be executed.
The determination of the fractal dimension itself was the next and final step. According to Eq.\ (\ref{gobs-ast_i}) if the galaxy distribution really formed a fractal system, the graphs of $\gamma_i^\ast$ vs.\ $d_i$ would behave as decaying power-law curves and the fractal dimension $D_i$ of the distribution would be determined by the linear fit slopes in the log-log plots.

\subsection{Results}
\subsubsection{Reduced subsample}
\begin{figure}[H]
\centering
	\includegraphics[width=180mm]{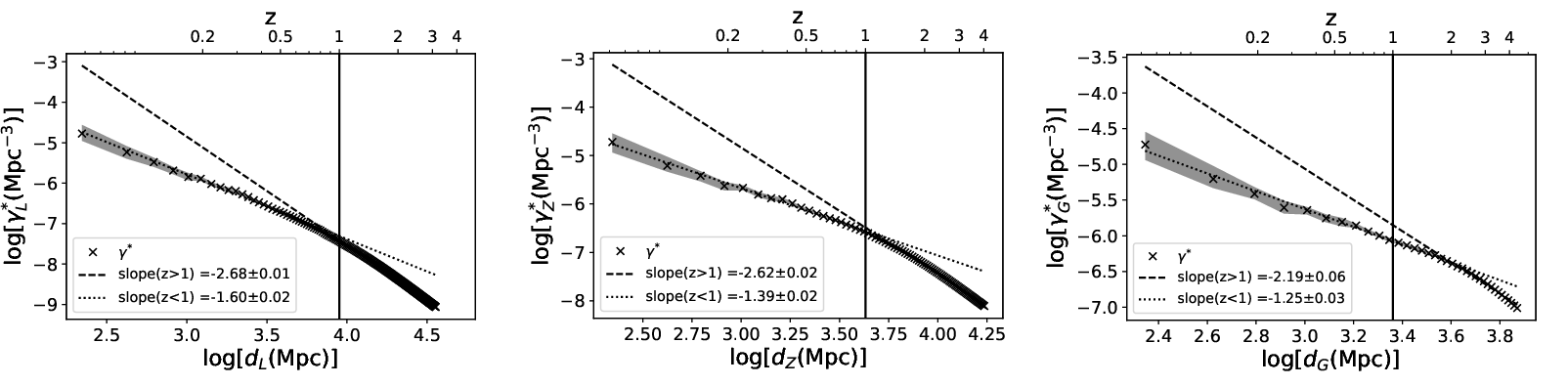}
    \caption[Log-log plot of $\gamma^*_{i}$ vs. $d_i$ in the UltraVISTA DR1 selected sample]{Log-log plots of $\gamma^*_{L}$ vs. $d_L$, $\gamma^*_{Z}$ vs. $d_Z$ and $\gamma^*_{G}$ vs. $d_G$, from left to right, obtained with the UltraVISTA DR1 galaxy redshift survey dataset. The dotted line is the linear fit for galaxies at $z<1$, while the dashed line represents those with $1\leq z\leq 4$. The grey area illustrates the 1 sigma error of the bins.}
    \label{fig:uvistaz4}
\end{figure}

The reduced UltraVISTA DR1 galaxy survey subsample characterized according to the criteria set at Sec. 3.1.1 agreed, to a good approximation, with the predictions that the galaxy distribution does form a fractal system as seen in graphs for log-log values of $\gamma_i^\ast$ vs.\ $d_i$. In addition, two power-law decaying regions were noted in the data, for $z<1$ and $z>1$, generating different single fractal dimensions. The results for all distance definitions adopted here are in Fig.\ \ref{fig:uvistaz4}, from where it can be determined that for $z<1$ the fractal dimension is in the range $1.38-1.78$, while for $1\le z\le4$ the resulting range is fairly smaller, $0.31-0.87$. Table \ref{tab1} assemble these results.

\begin{table}[H]
\caption[Fractal dimensions for the UltraVISTA DR1 reduced subsample]{Results for the UltraVISTA DR1 galaxy survey fractal analysis in two redshift scales in the \textit{reduced} subsample. The single fractal dimensions $D_{\ssty L}$, $D_{\ssty z}$ and $D_{\ssty G}$ were acquired from this galaxy distribution respectively using the luminosity distance $\dl$, redshift distance $\dz$ and galaxy area distance (transverse comoving distance) $\dg$.}
\label{tab1}
\begin{center}
\begin{tabular}{cccc}
\hline
& $D_{\ssty L}$ & $D_{\ssty z}$ & $D_{\ssty G}$\\
\hline
$z<1$         & $1.40\pm0.02$ & $1.61\pm0.02$ & $1.75\pm0.03$\\
$1\le z\le 4$ & $0.32\pm0.01$ & $0.38\pm0.02$ & $0.81\pm0.06$\\
\hline
\end{tabular}
\end{center}
\end{table}

These results indicate, in summary, that the UltraVISTA DR1 galaxy survey presents a galaxy distribution dataset that can be interpreted as a fractal system with the following approximate single fractal dimensions values: $D\,(z<1)=(1.58 \pm0.20)$ and $D\,(1\le z\le4)=(0.59\pm0.28)$. The potential reasons to why the fractal dimension is so much reduced at the deepest range $z>1$ will be discussed later.

\subsubsection{Complete (unselected) survey data}

The fractal analysis of the UltraVISTA DR1 galaxy subsample was established by a plot of absolute magnitudes in the $J$-band, as defined in Sec. 3.1.1, in terms of galaxies measured redshifts, as shown in Fig.\ \ref{fig:uvistaselection}. However, this plot suggests that the whole sample may be volume-limited, since even galaxies outside the absolute magnitude cut are distributed along increasing redshift bins. So, it is pertinent to apply the same fractal methodology developed here to the whole survey data, not only to the subsample, in order to compare the results.

\begin{figure}[H]
\centering
	\includegraphics[width=180mm]{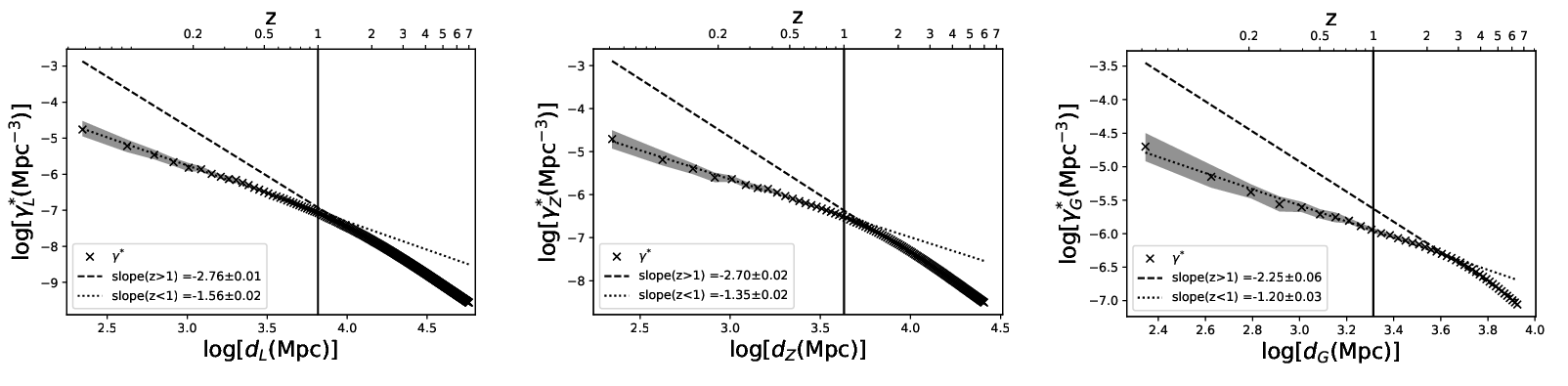}
    \caption[Log-log plot of $\gamma^*_{i}$ vs. $d_i$ in the UltraVISTA DR1 complete data]{Log-log plots of $\gamma^*_{L}$ vs. $d_L$, $\gamma^*_{Z}$ vs. $d_Z$ and $\gamma^*_{G}$ vs. $d_G$, from left to right, obtained with the UltraVISTA DR1 galaxy redshift survey dataset. The dotted line is the linear fit for galaxies at $z < 1$, while the dashed line represents those with $1\leq z\leq 7$. The grey area illustrates the 1 sigma error of the bins.}
    \label{fig:uvistaz7}
\end{figure}

The results for the fractal analysis of the complete UltraVISTA DR1 survey data are in Fig.\ \ref{fig:uvistaz7}. This dataset of galaxy distribution clearly presents fractal features in both regions, above and below the redshift value $z=1$. The respective single fractal dimensions for each distance definitions are in the range $1.42-1.83$ for $z<1$, while for $1\le z\le6$ the dimension is significantly smaller, in the range $0.23-0.81$. Table \ref{tab2} assembles these results and show that they are very much alike with the values from Table \ref{tab1} for the reduced subsample, although they do not overlap even considering their respective uncertainties.
So, it is evident that the whole UltraVISTA DR1 galaxy survey can also be described as a fractal system having two consecutive single fractal dimensions: $D\,(z<1) = (1.63\pm0.20)$ and $D\,(1\le z\le6) = (0.52\pm0.29)$.

\begin{table}
\caption[Fractal dimensions for the UltraVISTA DR1 complete subsample]{Results for the UltraVISTA DR1 galaxy survey fractal analysis in two redshift scales in the \textit{complete} subsample. The single fractal dimensions $D_{\ssty L}$, $D_{\ssty z}$ and $D_{\ssty G}$ were acquired from this galaxy distribution respectively using the luminosity distance $\dl$, redshift distance $\dz$ and galaxy area distance (transverse comoving distance) $\dg$.}
\label{tab2}
\begin{center}
\begin{tabular}{cccc}
\hline
& $D_{\ssty L}$ & $D_{\ssty z}$ & $D_{\ssty G}$\\
\hline
$z<1$         & $1.44\pm0.02$ & $1.65\pm0.02$ & $1.80\pm0.03$\\
$1\le z\le 6$ & $0.24\pm0.01$ & $0.30\pm0.02$ & $0.75\pm0.06$\\
\hline
\end{tabular}
\end{center}
\end{table}

\section{COSMOS2015 \& SPLASH}
\subsection{Data Selection}

\begin{figure}[H]
\centering
	\includegraphics[width=180mm]{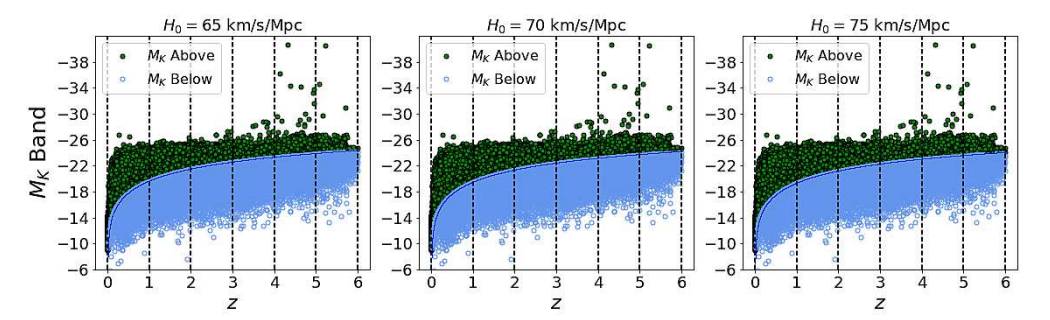}
    \caption[COSMOS2015 absolute magnitude selection]{Plot of the absolute magnitudes for the COSMOS2015 galaxies in terms of their photometrically measured redshift values. The dividing line corresponds to apparent magnitude $K_{\mathrm{S}} = 24.7$ and all galaxies having $M_K$ above this cutoff and $z \leq 4$ were included in the subsample that assumed three different values of the Hubble constant within its uncertainty.}
    \label{fig:cosmos2015selection}
\end{figure}

As stated above, fractal analysis of galaxy surveys demands arranging data along volume-limited distributions. Due to the fact that galaxy surveys are limited by apparent magnitude, we need to continue by reducing the data into subsamples so that they follow increasing redshift bins. This is done the same way as in Sec. 3.1.1 using Eq.(\ref{magabs1}) and assuming the FLRW cosmology with $\Omega_{m_{\ssty 0}}=0.3$, $\Omega_{\Lambda_{\ssty 0}}=0.7$ and $H_0=(70\pm5)\;\mbox{km}\; {\mbox{s}}^{-1} \; {\mbox{Mpc}}^{-1}$.
A homogeneous volume-limited subsample could be created by assuming the apparent magnitude threshold in the $K$ band to be $K=24.7$, which is a mean limiting value acceptable to both surveys \cite{teles2020}, and by changing Eq.\ (\ref{magabs1}) to the expression below,
\be
M_K=24.7-5\log \dl (z) -25.
\lb{magabs3}
\ee
From this expression we get the cutoff line between the selected and unselected galaxies. Finally, we took into consideration the uncertainty in the Hubble constant in order to test whether our results would be affected by its error margin.
Fig.\ \ref{fig:cosmos2015selection} shows the data selection of the COSMOS2015 survey obtained according to the method described previously. Only galaxies with absolute magnitudes $M_K$ above the cutoff line were selected as part of the subsample for further analysis. In addition, since galaxies having $z>4$ are in an amount too small to be considered representative, as we can see in Fig.\ \ref{fig:estatisticacatalogos}, we disregarded them. The outcome of this process was the creation of three subsamples containing 230705 galaxies in the redshift range $0 < z\leq4$ from the original 578379 objects for three values of the Hubble constant within its uncertainty.

\begin{figure}[H]
\centering
	\includegraphics[width=180mm]{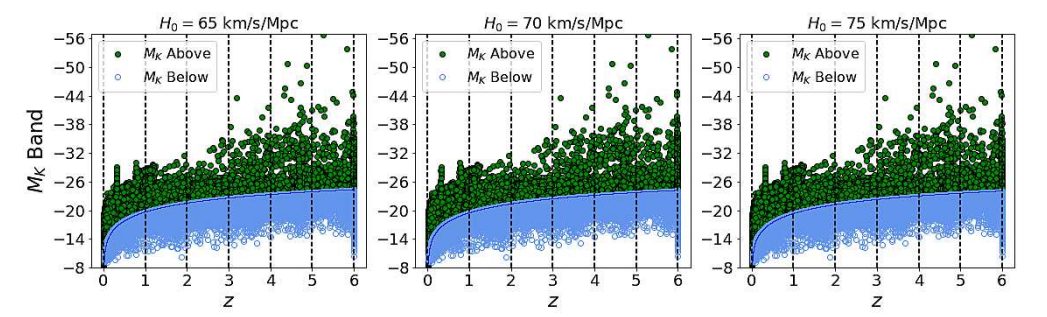}
    \caption[SPLASH absolute magnitude selection]{Plot of the absolute magnitudes for the SPLASH galaxies in terms of their photometrically measured redshift values. The dividing line corresponds to apparent magnitude $K = 24.7$ and all galaxies having $M_K$ above this cutoff and $z \leq 4$ were included in the subsample that assumed three different values of the Hubble constant within its uncertainty.}
    \label{fig:splashselection}
\end{figure}

Fig.\ \ref{fig:splashselection} shows the same selection procedure executed with the SPLASH survey dataset also generating three other subsamples. The original 390362 objects were then reduced to 171548 galaxies.

\subsection{Data Analysis}

The FLRW cosmology parameters allowed the calculation of the observational distances $d_i\,(i={\ssty G}$, ${\ssty L}$, ${\sty z})$ using theoretical expressions relating distance to redshift given by this cosmological model. Together with the photometric redshift values supplied by the galaxy surveys these observational distances can be calculated. The data sorting algorithm necessary to implement a fractal analysis under the theoretical model discussed in Sec. 1.2.2 is the same as described in Sec. 3.1.2. 
For each subsample, we start by setting the minimum redshift $z_{\ssty 0}$, its respective minimum distance $d_{i_0}=d_{i_0}(z_{\ssty 0})$, and the incremental distance interval $\Delta d_i$. The value $\Delta d_i=200$~Mpc was kept in all cases, as provides a large amount of data elements for all quantities involved, allowing enough points to perform an adequate regression analysis. Then, the algorithm establishes the first bin by counting the number of galaxies $N_{i_1}$ within $d_{i_0}$ and $d_{i_1}=d_{i_0}+\Delta d_i$, and calculating its respective volume density $\gamma_{i_1}^\ast$. The next step is to increase the bin size by $\Delta d_i$, leading to a distance of $d_{i_2}=d_{i_0}+2\Delta d_i$. At a interval between $d_{i_0}$ and $d_{i_2}$, we count the number of galaxies $N_{i_2}$ and estimate $\gamma_{i_2}^\ast$. These steps are repeated $n$ times until all galaxies are included.
Finally, if the galaxy distribution truly formed a fractal system, the behavior of $\gamma_i^\ast$ vs.\ $d_i$ plots would be as decaying power-law curves according to Eq.\ (\ref{gobs-ast_i}). So the linear fit slopes in log-log plots would make it possible to directly determine the fractal dimensions $D_i$ of the distribution.

\subsection{Results}

\begin{figure}[H]
\centering
	\includegraphics[width=180mm]{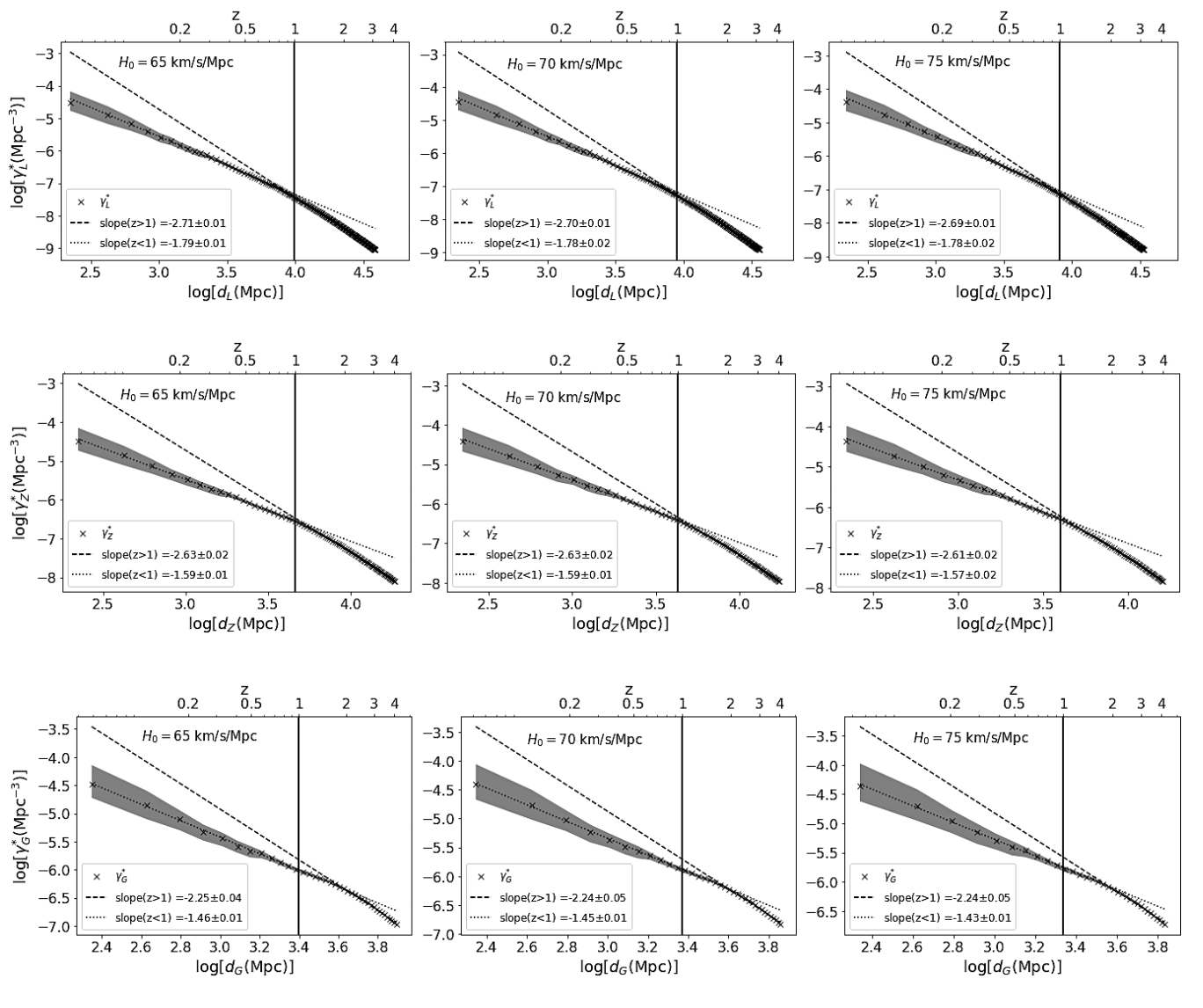}
    \caption[Log-log plot of $\gamma^*_{i}$ vs. $d_i$ for the COSMOS2015 survey]{Log-log plots of $\gamma^*_{L}$ vs. $d_L$ (top), $\gamma^*_{Z}$ vs. $d_Z$ (middle) and $\gamma^*_{G}$ vs. $d_G$ (bottom), with three different values of the Hubble constant, obtained with the COSMOS2015 galaxy redshift survey dataset. The dotted line is the linear fit for galaxies at $z < 1$, while the dashed line represents those with $1\leq z\leq 4$. The grey area illustrates the 1 sigma error of the bins.}
    \label{fig:cosmos2015gradez4}
\end{figure}

\begin{figure}[H]
\centering
	\includegraphics[width=180mm]{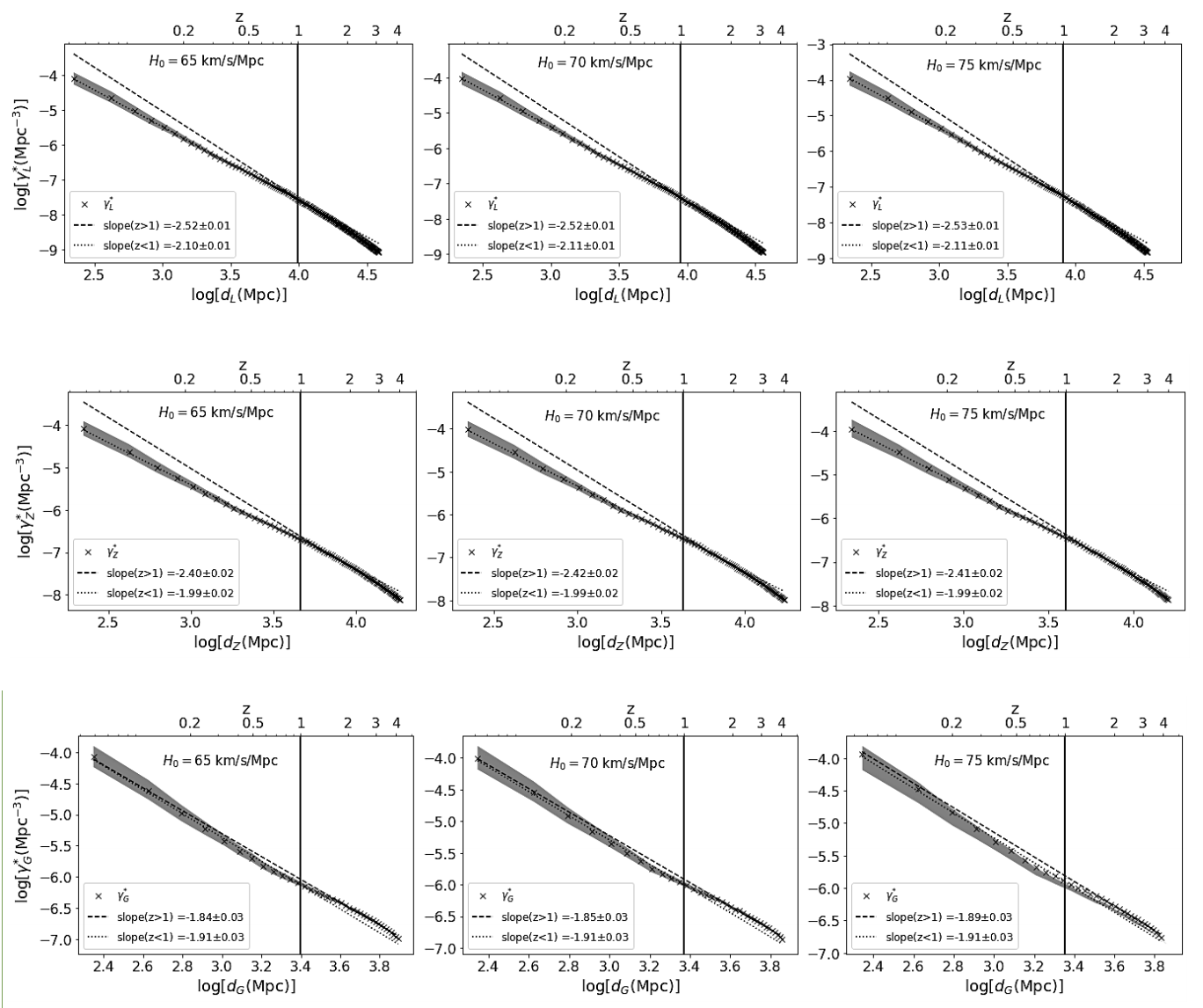}
    \caption[Log-log plot of $\gamma^*_{i}$ vs. $d_i$ for the SPLASH survey]{Log-log plots of $\gamma^*_{L}$ vs. $d_L$ (top), $\gamma^*_{Z}$ vs. $d_Z$ (middle) and $\gamma^*_{G}$ vs. $d_G$ (bottom), with three different values of the Hubble constant, obtained with the SPLASH galaxy redshift survey dataset. The dotted line is the linear fit for galaxies at $z < 1$, while the dashed line represents those with $1\leq z\leq 4$. The grey area illustrates the 1 sigma error of the bins.}
    \label{fig:splashgradez4}
\end{figure}

Figs.\ \ref{fig:cosmos2015gradez4} and \ref{fig:splashgradez4} show a grid of the $\gamma_i^\ast$ vs.\ $d_i$ log-log graphs with their respective Hubble constant values choices for the COSMOS2015 and SPLASH surveys. The results reveal that the galaxy distribution in both surveys display power-law decays in two scale ranges: for $z<1$ and $1\leq z\leq 4$, which is coherent with a fractal system presenting two single fractal dimensions at different distance ranges.
Using Eq.\ (\ref{slope3}) we can simply calculate the fractal dimension in both scaling ranges from the slopes of the fitted straight lines. These results obtained with this procedure are shown in Tables \ref{tabcosmos} and \ref{tabsplash}, where one finds two single fractal systems in sequence at different data ranges with decreasing values for $D$ as the redshift increases. Hence, the results confirm the theoretically predicted decay in the fractal dimension the deeper one goes in the surveys (see above in the Introduction). In addition, the fractal dimensions are unaffected by variations in the Hubble constant as demonstrate by the estimated values being within 1 sigma error in both COSMOS2015 (Table \ref{tabcosmos}) and SPLASH (Table \ref{tabsplash}).
Summarizing the results of Tables \ref{tabcosmos} and \ref{tabsplash} by rounding off the main values and uncertainties around their medians, we can determine that for $z<1$ the COSMOS2015 survey presented $D=1.4\pm0.2$, while the SPLASH galaxies generated $D=1.0\pm0.1$. For $1\leq z\leq4$ we respectively found $D=0.5\pm0.3$ and $D=0.8\pm0.4$. Evidently the SPLASH galaxies obtained fractal dimensions somewhat smaller than the COSMOS2015 ones for $z<1$, but the opposite was true for $z>1$, even though with overlapping uncertainties. 

Next, there will be a discussion for the potential reasons for such differences and a comparison with previous studies.
Finally, Table \ref{tabcompara} contain the comparison of the fractal dimensions $D$ calculated with the UltraVISTA DR1, COSMO2015, SPLASH and FDF (see Ref. \cite{gabriela}) surveys when adopting $H_0=70$ km/s/Mpc. 

\begin{table}[H]
\small
\caption[Fractal dimensions for COSMOS2015 varying the Hubble constant $H_0$]{Fractal dimensions calculated in the selected subsamples of the COSMOS2015 galaxy redshift survey in the ranges $z<1$ and $1\leq z\leq4$. The single fractal dimensions $D_{L}$, $D_{Z}$ and $D_{G}$ were obtained from the galaxy distributions respectively using the luminosity distance $d_L$, redshift distance $d_Z$ and galaxy area distance (transverse comoving distance) $d_G$.}
\label{tabcosmos}
\begin{center}
\begin{tabular}{ccccccc}
\hline
$H_0 $ & $D_L (z < 1)$ & $D_Z (z<1)$ & $D_G (z<1)$ & $D_L (z>1)$ & $D_Z (z>1)$ & $D_G (z>1)$ \\
\hline
$65$ km/s/Mpc  & $1.21\pm0.01$ & $1.41\pm0.01$ & $1.54\pm0.01$ & $0.29\pm0.01$ & $0.37\pm0.02$ & $0.75\pm0.04$\\

$70$ km/s/Mpc & $1.22\pm0.02$ & $1.41\pm0.01$ & $1.55\pm0.01$& $0.30\pm0.01$ & $0.37\pm0.02$ & $0.76\pm0.05$\\

$75$ km/s/Mpc & $1.22\pm0.02$ & $1.43\pm0.02$ & $1.57\pm0.01$& $0.31\pm0.01$ & $0.39\pm0.02$ & $0.76\pm0.05$\\
\hline
\end{tabular}
\end{center}
\end{table}

\begin{table}[H]
\small
\caption[Fractal dimensions for SPLASH varying the Hubble constant $H_0$]{Fractal dimensions calculated in the selected subsamples of the SPLASH galaxy redshift survey in the ranges $z<1$ and $1\leq z\leq4$. The single fractal dimensions $D_{L}$, $D_{Z}$ and $D_{G}$ were obtained from the galaxy distributions respectively using the luminosity distance $d_L$, redshift distance $d_Z$ and galaxy area distance (transverse comoving distance) $d_G$.}
\label{tabsplash}
\begin{center}
\begin{tabular}{ccccccc}
\hline
$H_0 $ & $D_L (z < 1)$ & $D_Z (z<1)$ & $D_G (z<1)$ & $D_L (z>1)$ & $D_Z (z>1)$ & $D_G (z>1)$ \\
\hline
$65$ km/s/Mpc  & $0.90\pm0.01$ & $1.01\pm0.02$ & $1.09\pm0.03$& $0.48\pm0.01$ & $0.60\pm0.02$ & $1.16\pm0.03$\\

$70$ km/s/Mpc & $0.89\pm0.01$ & $1.01\pm0.02$ & $1.09\pm0.03$& $0.48\pm0.01$ & $0.58\pm0.02$ & $1.15\pm0.03$\\

$75$ km/s/Mpc & $0.89\pm0.01$ & $1.01\pm0.02$ & $1.09\pm0.03$& $0.47\pm0.01$ & $0.59\pm0.02$ & $1.11\pm0.03$\\
\hline
\end{tabular}
\end{center}
\end{table}

\begin{table}[H]
\scriptsize
\caption[Fractal dimension comparison]{The comparison displayed of all recently calculated single fractal dimensions applying similar analytical tools as presented here to various galaxy distribution surveys. These results were obtained with the UltraVISTA DR1 \cite{ultra1}, COSMOS2015 \cite{La2016} and SPLASH \cite{Mehta2018}, and FDF \cite{fdf2004, fdf2006} catalogs, all considering $H_0=70$ km/s/Mpc. There is a clear tendency for decreasing values of $D$ at $z>1$ in virtually all results, as theoretically predicted in the Introduction. Such a decrease is, nonetheless, less pronounced in the SPLASH data, which is the only galaxy distribution shown here to have been surveyed in the southern hemisphere.}
\label{tabcompara}
\begin{center}
\begin{tabular}{ccccccccc}
\hline
& UVista DR1 & $(0.2<z<4)$ & COSMO2015 & $(0.1<z<4)$ &
SPLASH & $(0.1<z<4)$ & FDF & $(0.45<z<5)$ \\
 & $z<1.0$ & $z>1.0$ & $z<1.0$ & $z>1.0$ & $z<1.0$ & $z>1.0$ &
 $z\lesssim1.2$ & $z\gtrsim1.2$ \\
\hline
$D_{\ssty L}$ & $1.40\pm0.02$ & $0.32\pm0.01$ & $1.22\pm0.02$ & $0.30\pm0.01$ &
   $0.89\pm0.01$ & $0.48\pm0.01$ & $1.2\pm0.3$ & $0.5\pm0.2$ \\

$D_{\ssty z}$ & $1.61\pm0.02$ & $0.38\pm0.02$ & $1.41\pm0.01$ & $0.37\pm0.02$ &
   $1.01\pm0.02$ & $0.58\pm0.02$ & $1.5\pm0.4$ & $0.6\pm0.2$ \\

$D_{\ssty G}$ & $1.75\pm0.03$ & $0.81\pm0.06$ & $1.55\pm0.01$ & $0.76\pm0.05$ &
   $1.09\pm0.03$ & $1.15\pm0.03$ & $1.8\pm0.3$ & $1.0\pm0.7$ \\
\hline
\end{tabular}
\end{center}
\end{table}

\subsection{Blue and red subsamples}

A question that can be raised regarding this fractal analysis is the possibility that the single fractal dimension $D$ would depend on galactic type, features or potential evolutionary stages, and if there are dependencies on some, or all, of these characteristics. In such a case then the fractal dimension could, perhaps, be used as a tracer of galactic properties or their evolutionary stages. In this subsection we introduce a simple test of this possible dependency using the COSMOS2015 and SPLASH surveys studied here. 

\begin{figure}[H]
\centering
\includegraphics[width=100mm]
{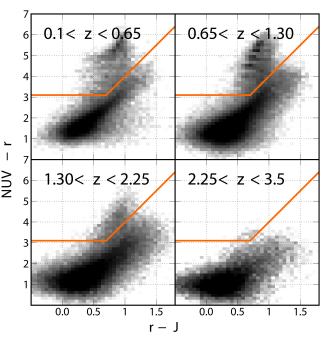}
\caption[NUV - \textit{r/r} - \textit{J} COSMOS2015 galaxy distributions]{NUV - \textit{r/r} - \textit{J} COSMOS2015 galaxy distributions. Quiescent galaxies lie in the top-left corner. The objects fainter than limiting magnitudes are not considered \cite{La2016}.}
\label{fig:cosmos2015_selection_type}
\end{figure}

These two surveys allow us to compute $D$ in two galaxy subsamples: the red, or quiescent, and the blue, or star forming galaxies. Therefore, this grants us with a selection that provides an initial and direct way for testing the concept of using $D$ as a tracer of galactic characteristics. The criteria for creating these subsamples uses color-color diagrams or star formation rates as supplied in both surveys databases.
Regarding the COSMOS2015 dataset, the classification introduced in Ref. \cite{La2016} is obtained from the location of galaxies in the NUV-r vs.\ r-J color-color diagram \cite{Wi2009}. In addition to using these colors, this selection has the absolute magnitudes' estimation at rest-frame based on the apparent magnitude at $\lambda_{\mbox{\footnotesize rest-frame}}(1+z_{\mbox{\footnotesize gal}})$, which reduces the k-correction dependency \cite{Il2005}. Such approach prevents the mixing between red dusty galaxies and quiescent ones. In implementation, selected quiescent galaxies have ${M}_{\mathrm{\ssty NUV}}-{M}_{r}> 3({M}_{r}-{M}_{J})+1$ and ${M}_{\mathrm{\ssty NUV}}-{M}_{r}> 3.1$, while the other galaxies were considered star-forming \cite{ultra1}. The redshift dependent evolution of this distribution is in Fig. \ref{fig:cosmos2015_selection_type}, where there is an accumulation of quiescent galaxies at low redshift inside the box and a relative decrease in bright, star-forming galaxies outside the box.

For the SPLASH survey, following Refs. \cite{Il2010} and \cite{DS2011} we divided the galaxy sample using the \textit{specific star formation rate} (SSFR). Galaxies with $\log \mbox{SSFR}>-11$ were classified as star-forming, whereas the ones with $\log \mbox{SSFR}<-11$ were set as quiescent. SSFR is the ratio between star forming rate and stellar mass, being a simple approach for quantifying in a uniform way the star formation degree. Such separation by stellar mass indicates how strong is the star formation in a certain galaxy, which is particularly important when comparing galaxies having different masses and sizes. In Ref. \cite{ultra11} this classification was compared to the color-color selection applied to COSMOS2015, and the conclusion was that both methods provide similar results at $z<1$, but towards high redshift the one based on SSFR tends to be more conservative.

\begin{figure}[H]
\centering
\includegraphics[width=140mm]{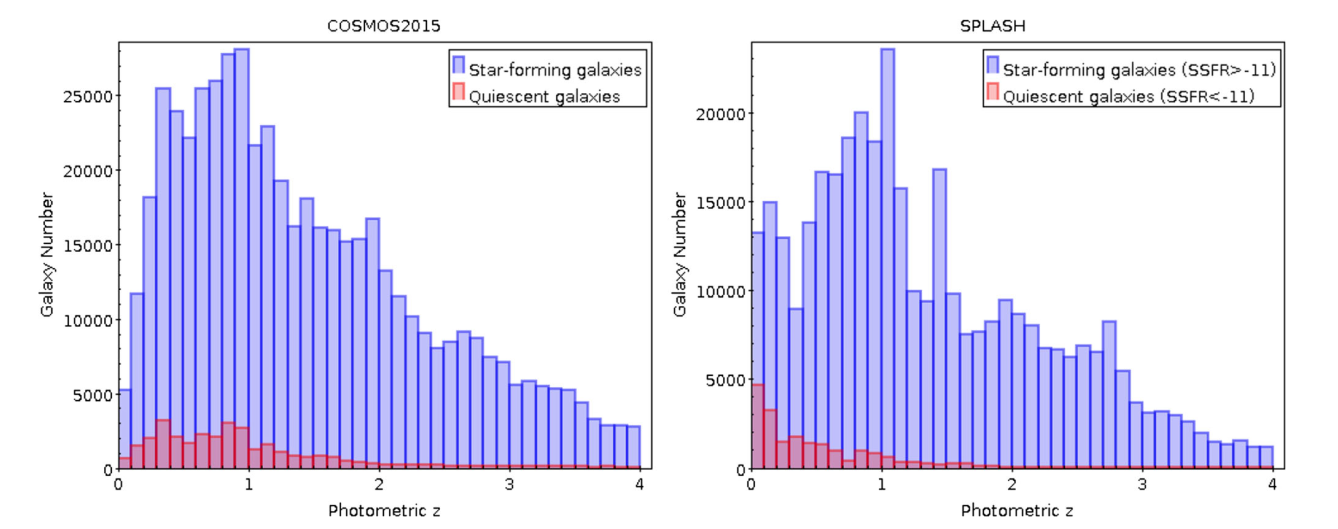}
 \caption[Histogram for blue and red galaxies of COSMOS2015 and SPLASH]{Histograms displaying the galaxy distribution numbers in terms of the redshift for the COSMOS2015 (left) and SPLASH (right) subsamples of blue star forming galaxies and red quiescent ones. Labels are as in the legends. COSMOS2015 galaxies were classified by color and SPLASH galaxies were separated considering the specific stellar formation rate (SSFR) using -11 as the cutoff value (see main text). Clearly the number of blue galaxies is much higher than the red ones in both surveys.} 
  \label{blue-red-cosmos-splash-classes}
\end{figure}

\begin{figure}[H]
\centering
\includegraphics[width=130mm]{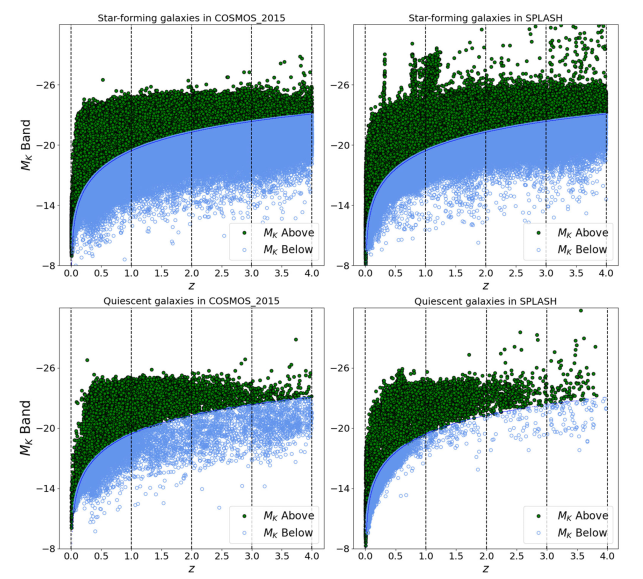}
\caption[Absolute magnitudes for blue and red galaxies of COSMOS2015 and SPLASH]{Plots of the absolute magnitudes for the blue star forming galaxies (top), and red quiescent ones (bottom) for the COSMOS2015 (left) and SPLASH (right) surveys in terms of their photometrically measured redshift values. Specifications are as in the title of each plot. As in Figs.\ \ref{fig:cosmos2015selection} and \ref{fig:splashselection}, the dividing line corresponds to apparent magnitude $K = 24.7$, so only blue and red galaxies having $M_K$ above this cutoff and $z\leq 4$ were included in the blue and red COSMOS2015 and SPLASH subsamples. Since previous results indicate that the fractal dimension is not affected by altering the Hubble constant within its currently accepted uncertainty, here only $H_0=70$ km/s/Mpc was assumed.}
  \label{mag_complete-blue-red-cosmos-splash}
\end{figure}

Two subsamples were generated for each survey taking into consideration the criteria explained above. Fig.\ \ref{blue-red-cosmos-splash-classes} displays histograms of redshift distribution of both subsamples up to $z=4$, where it is evident that the number of blue star forming galaxies is greater than red quiescent ones. This selection methodology led the COSMOS2015 survey to turn out as 527899 star forming galaxies and 31424 quiescent ones. The SPLASH was respectively divided into 359021 and 20045 galaxies. Then, the same volume-limited filtering process of obsolute magnitude cutoff was applied to these four subsamples as defined by Eq.\ (\ref{magabs3}). In Fig.\ \ref{mag_complete-blue-red-cosmos-splash} is presented the outcome of this filtering procedure, which filtered the COSMOS2015 subsamples to 208005 blue galaxies and 22824 red ones, while reducing the SPLASH data to 205012 and 13491 galaxies respectively.


\begin{figure}[H]
\centering
 \includegraphics[width=180mm]{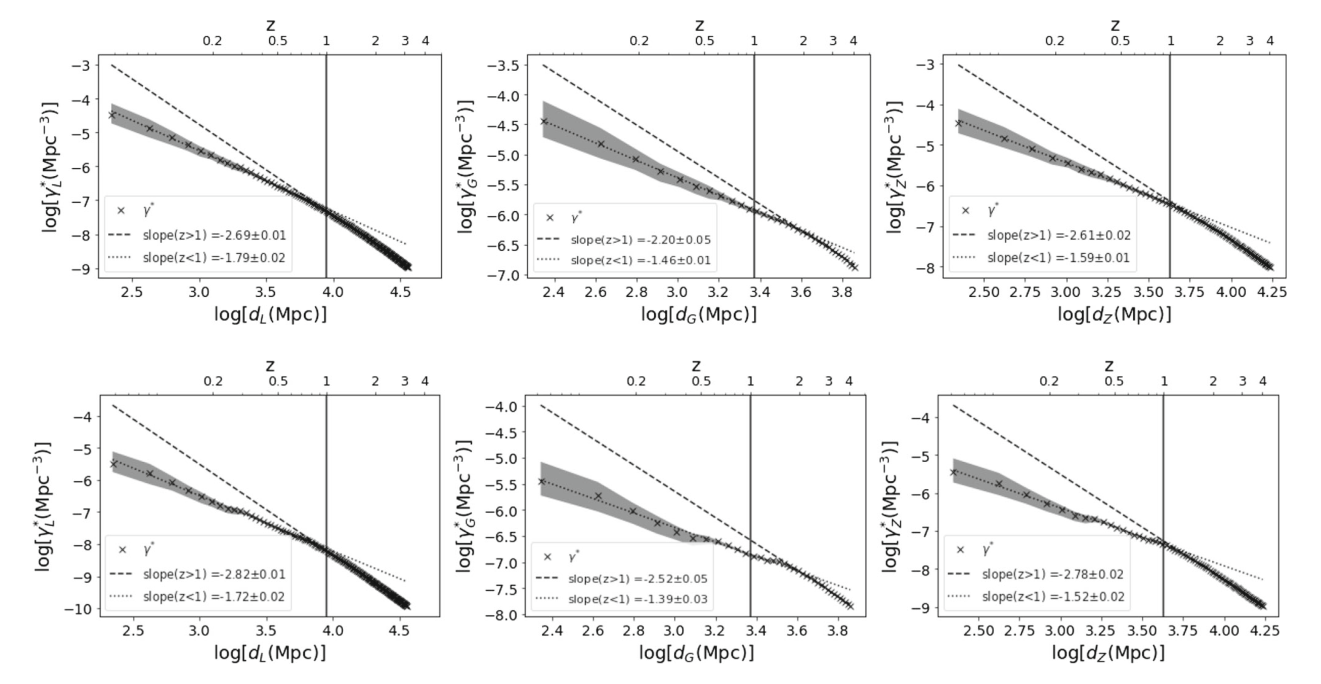}
 \caption[Log-log plot of $\gamma^*_{i}$ vs. $d_i$ for the blue and red COSMOS2015 galaxies]{Log-log graph of $\gaml^\ast$, $\gamg^\ast$ and $\gamz^\ast$ respectively vs.\ $\dl$, $\dg$ and $\dz$ obtained with the blue COSMOS2015 star forming galaxy subsample (top) and with the red COSMOS2015 quiescent galaxy subsample (bottom). The dotted line is the linear fit for galaxies at $z<1$, while the dashed line represents those with and $1\leq z\leq 4$. The grey area illustrates the 1 sigma error of the bins. All results were obtained considering $H_0=70$ km/s/Mpc.}
\label{blue-red-cosmosdLdGdz}
\end{figure}

\begin{figure}[H]
\centering
 \includegraphics[width=180mm]{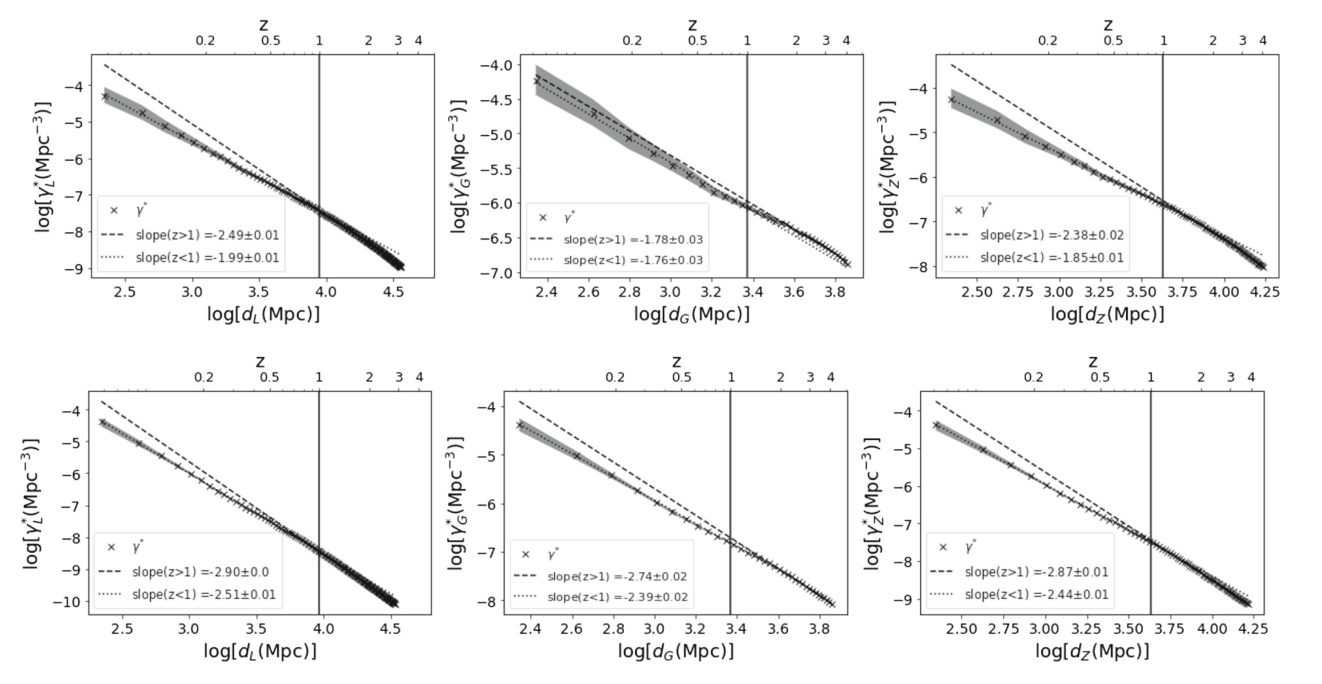}
 \caption[Log-log plot of $\gamma^*_{i}$ vs. $d_i$ for the blue and red SPLASH galaxies]{Log-log graph of $\gaml^\ast$, $\gamg^\ast$ and $\gamz^\ast$ respectively vs.\ $\dl$, $\dg$ and $\dz$ obtained with the blue SPLASH star forming galaxy subsample (top) and with the red SPLASH quiescent galaxy subsample (bottom). The dotted line is the linear fit for galaxies at $z<1$, while the dashed line represents those with and $1\leq z\leq 4$. The grey area illustrates the 1 sigma error of the bins. All results were obtained considering $H_0=70$ km/s/Mpc.}
\label{blue-red-splashdLdGdz}
\end{figure}

The calculation method for the $\gamma_i^\ast$ number densities from Sec.\ 3.2.2 was adopted for these four selected and filtered galaxies subsamples, then the resulting data points were linear fitted against their respective distance measurement values $d_i\,(i={\ssty G}$, ${\ssty L}$, ${\sty z})$. Figs.\ \ref{blue-red-cosmosdLdGdz} and \ref{blue-red-splashdLdGdz} show the decaying power-law curves and data fits in the ranges $z<1$ and $1\leq z\leq 4$. Considering the robust results for the fractal dimension against changes in the Hubble constant as presented in Tables \ref{tabcosmos} and \ref{tabsplash}, in this subsection we calculated all results using only $H_0=70$ km/s/Mpc.


The fractal dimensions obtained from the graphs in Figs.\ \ref{blue-red-cosmosdLdGdz} and \ref{blue-red-splashdLdGdz} are assembled in Table~\ref{tabA}, which presents results that can now be compared with the 3rd to 6th columns of Table \ref{tabcompara}. The blue COSMOS2015 galaxies had their fractal dimensions essentially unchanged as compared to the unselected samples, while the blue SPLASH ones had $D$ moderately increased for $z<1$, but remained basically unchanged for $z>1$ within the uncertainties. However, the fractal dimensions of the quiescent galaxies had more evident changes. For $z<1$ the red COSMOS2015 subsample had higher values for $D$ in all distance measures, but experienced a fractal dimension reduction at $z>1$. The same substantial decrease in $D$ happened for the red SPLASH galaxies in both ranges, $z<1$ and $z>1$. Finally, in all cases for every distance measures the fractal dimension reduction in the range $z>1$ is observed, as the theory predicted.

\begin{sidewaystable}
    \caption[Fractal dimensions for the blue and red galaxy subsamples]{Fractal dimensions determined in the selected blue-star-forming and red-quiescent, and then volume-limit filtered, galaxy subsamples of the COSMOS2015 and SPLASH redshift surveys in the range $z\leq 4$. The single fractal dimensions $D_{\ssty L}$, $D_{\ssty z}$ and $D_{\ssty G}$ were acquired from the galaxy distributions respectively using the luminosity distance $\dl$, redshift distance $\dz$ and galaxy area distance (transverse comoving distance) $\dg$. The results were calculated considering only $H_0=70$ km/s/Mpc (see main text). A comparison of these numbers with the ones in the 3rd to 6th columns of Table \ref{tabcompara} shows that the fractal dimensions do vary according to the blue-red selection used here, this being especially the case for the quiescent galaxies. Similarly, as in the results shown in Table \ref{tabcompara}, all values of $D$ decrease in the range $z>1$, in some cases quite substantially.}
\label{tabA}
\begin{center}
\begin{tabular}{ccccccccc}
\hline
& Blue COSMOS2015 & $\sty (0.1<z<4)$ &
  Blue SPLASH & $\sty (0.1<z<4)$ &
  Red COSMOS2015 & \scalebox{0.8}{$(0.1<z<4)$} &
  Red SPLASH & $\sty (0.1<z<4) $ \\

& $z<1.0$ & $z>1.0$ & $z<1.0$ & $z>1.0$ & $z<1.0$ &
$z>1.0$ & $z<1.0$ & $z>1.0$ \\
\hline
$D_{\ssty L}$ & $1.21\pm0.02$ & $0.31\pm0.01$ & $1.01\pm0.01$ & $0.51\pm0.01$ &
   $1.28\pm0.02$ & $0.18\pm0.01$ & $0.49\pm0.01$ & $0.10\pm0.01$ \\

$D_{\ssty z}$ & $1.41\pm0.01$ & $0.39\pm0.02$ & $1.15\pm0.01$ & $0.62\pm0.02$ &
   $1.48\pm0.02$ & $0.22\pm0.02$ & $0.56\pm0.01$ & $0.13\pm0.01$ \\

$D_{\ssty G}$ & $1.54\pm0.01$ & $0.80\pm0.05$ & $1.24\pm0.03$ & $1.22\pm0.03$ &
   $1.61\pm0.03$ & $0.48\pm0.05$ & $0.61\pm0.02$ & $0.26\pm0.02$ \\
\hline
\end{tabular}
\end{center}
\end{sidewaystable}

\chapter*{Conclusion}
\addcontentsline{toc}{chapter}{Conclusion}
This dissertation pursued to empirically test if the large-scale galaxy distribution could be represented as a fractal system and if galaxy types could possibly be dependent on the single fractal dimension $D$.
Some of the Newtonian hierarchical cosmology relations were extended and applied to relativistic cosmological models in order to describe possible galaxy fractal structures by means of $D$ at deep redshift scales.
The application of these tools to the UltraVISTA DR1, COSMOS2015 and SPLASH datasets comprised almost one million objects spanning the redshift interval of $0.1 \leq z \leq 6$.

Absolute magnitudes were calculated using the measured redshifts of the surveys in the interest of acquiring volume-limited subsamples. These absolute magnitudes were adopted to obtain the luminosity distances $d_L$ by assuming the $\Lambda$CDM relativistic cosmological model and the apparent magnitude limit of 24 in the J-band for the UltraVISTA DR1 and 24.7 in the K-band for the COSMOS2015 and SPLASH, as well as data redshift cutoff at z=4. Since these subsamples showed that their galaxies followed increasing redshift bins, they were considered as effectively forming a volume-limited distribution. This procedure provided three subsamples with about 568k objects, the first containing 166566 UltraVISTA DR1 galaxies, the second containing 230705 COSMOS2015 galaxies, and the third containing 171548 SPLASH ones. Fractal analysis was then performed in these three subsamples. 

As relativistic cosmologies have several definitions of observed distance \cite{ellis2007}, only three were used here, which are the luminosity distance $d_L$, the redshift distance $d_Z$ and the galaxy area distance (transverse comoving distance) $d_G$. The adoption of several cosmological distance measures originates from the fact that relativistic effects become strong enough for redshift ranges larger than $z \gtrsim 0.3$, so these distance definitions generate different results for the same redshift value at those ranges. 
A data sorting algorithm was executed to plot ghaphs for number densities vs. relativistic distances. Straight lines were then fitted to the data in two scales, $z<1$ and $1 \leq z \leq 4$, whose slopes directly produced the single fractal dimensions.

The results indicate that the subsamples have two consecutive redshift ranges behaving as single fractal structures. Rounding them off and their respective uncertainties we found that for $z<1$ the UltraVISTA DR1 galaxies presented $D = (1.6 \pm 0.2)$, the COSMOS2015 galaxies produced $D = (1.4 \pm 0.2)$ and the SPLASH galaxies generated $D = (1.0 \pm 0.1)$. For $1 \leq z \leq 4$ the dimensions respectively decreased to $D = (0.6 \pm 0.3)$, $D = (0.5 \pm 0.3)$ and $D = (0.8 \pm 0.4)$. These results were not affected by changes in the Hubble constant within the assumed uncertainties. Additionally, no transition to observational homogeneity was found in the data.

Subsamples were then created by selecting blue star forming galaxies and red quiescent ones from the COSMOS2015 and SPLASH datasets. After filtering the data by the same absolute magnitude limits that were applied to the unselected samples, the results showed that up to two decimal digits the fractal dimension of blue galaxies remained essentially unchanged, whereas some red galaxies showed evident reduction at the same precision. This demonstrates that the fractal dimensions of both surveys are dominated by blue galaxies. These results also suggest that the fractal dimensions can be valuable tools as tracers of galaxy types and evolution in addition to being descriptors of galaxy distributions. So, in this context the fractal dimension would be seen as an intrinsic property of the objects' distribution in the Universe that could be adjusted depending on the types of galaxy formed, which means that $D$ can be applied as a tool to study different galaxy populations.

Finally, the results introduced here raise the question of why there is such considerable decrease in the fractal dimension for redshift values larger than the unity. The reason could be an observational effect caused by data bias, because many galaxies are not detected beyond $z>1$ reducing then the observed galaxy clustering and consequently reducing the associated fractal dimension. 
Another potential reason is bias associated to the small angular areas of these surveys so that they would not provide representative measurements of the entire sky distribution. 
Apart from observational biases, it is possible that the decrease in the fractal dimension is due to real physical effects. Considering galaxy evolution and large-scale structure dynamics of the Universe, it is plausible that there might be much less galaxies at high $z$. These interpretation is in agreement with the theoretical prediction that states a sharp decrease in observed number density at $z>1$ as shown in Ref. \cite{ribeiro95} and Fig. 6 in Ref. \cite{Albani2007}. So, the shift in $D$ to smaller values could be the empirical verification of this theoretical prediction, meaning that the Universe was void dominated at those epochs as a result of galaxies being much more sparsely distributed and in reduced numbers.



\appendix
\chapter{Python code used in this study}
\renewcommand{\theequation}{\thechapter.\arabic{equation}}

\definecolor{dkgreen}{rgb}{0,0.6,0}
\definecolor{gray}{rgb}{0.5,0.5,0.5}
\definecolor{mauve}{rgb}{0.58,0,0.82}
 
\lstset{
  language=Python,                
  basicstyle=\footnotesize,           
  numbers=left,                   
  numberstyle=\tiny\color{gray},  
  stepnumber=2,                             
  numbersep=5pt,                  
  backgroundcolor=\color{white},    
  showspaces=false,               
  showstringspaces=false,         
  showtabs=false,                 
  frame=single,                   
  rulecolor=\color{white},        
  tabsize=2,                      
  captionpos=b,                   
  breaklines=true,                
  breakatwhitespace=false,        
  title=\lstname,                               
  keywordstyle=\color{blue},          
  commentstyle=\color{dkgreen},       
  stringstyle=\color{mauve},     
}


            \begin{lstlisting}

# Analysis of the COSMOS2015 survey for z<4

 # Code to perform fractal analysis in large astronomical surveys
import pandas as pd
import matplotlib.pyplot as plt
import numpy as np
import astropy.units as u
from scipy import constants
from scipy import stats
from scipy.stats import linregress
from scipy import odr
import matplotlib.ticker as ticker
from astropy.cosmology import LambdaCDM, z_at_value
paramet = LambdaCDM(H0=70, Om0=0.3, Ode0=0.7)

# Importing tables
tabela_qst = pd.read_csv("quiescent_cosmos2015.csv")
tabela_sf = pd.read_csv("sf_cosmos2015.csv")

# Selecting sample
tabela=[]
sel_uvista = (tabela_sf['FLAG_HJMCC']==0)&(tabela_sf['TYPE']==0)&(tabela_sf['ZPDF']>0.0)
tabela = tabela_sf[sel_uvista]
dl = paramet.luminosity_distance(tabela.ZPDF).values.value
selacima  = (tabela['MK'] <= (-5*np.log10(dl) - 0.3)) & (tabela['ZPDF'] < 4.0) & (tabela['MK'] > -90)
MK_ABOVE = tabela[selacima]

dl = paramet.luminosity_distance(MK_ABOVE.ZPDF).values.value

# Unity of H0 is km/Mpc.s
H0 = paramet.H(0)
c = 299792 #km/s
z = MK_ABOVE.ZPDF
dz = (c*z) / H0
dg = dl/(1+z)

# Inferior uncertainty
dl_inf = paramet.luminosity_distance(MK_ABOVE.ZPDF_L68).values.value
H0 = paramet.H(0)
c = 299792 #km/s
z_inf = MK_ABOVE.ZPDF_L68
dz_inf = (c*z_inf) / H0
dg_inf = dl_inf/(1+z_inf)

# Upper uncertainty
dl_sup= paramet.luminosity_distance(MK_ABOVE.ZPDF_H68).values.value
H0 = paramet.H(0)
c = 299792 #km/s
z_sup = MK_ABOVE.ZPDF_H68
dz_sup = (c*z_sup) / H0
dg_sup = dl_sup/(1+z_sup)

 # Functions to assist in our analysis
 # 1) Find redshift from dl, in order to find dg(z) relation
 def find_z(dist):
    z_test=np.arange(0.001,7.05,0.0001)
    dg_test = paramet.luminosity_distance(z_test).value/(1+z_test)
    i=0
    test= dg_test[i]
    while round(test,2)<round(dist,2):
        i=i+1
        test=dg_test[i]
        z_out=z_test[i]
        dist_out = test
    return z_out

# 2) Function to generate gama density in respect to a given distance
def gama_dist(dbin, dist, dist_sup, dist_inf, dist_type):
  list_count = []
  list_count_sup = []
  list_count_inf = []
  list_gama = []
  list_gama_sup = []
  list_gama_inf = []
  list_dist = []
  list_z = []
  count = 0
  dmin = min(dist)

  while count < len(dist):
    d = dmin + dbin
    # Center values:
    sel = dist<(d)
    dist_sel=dist[sel]
    count = len(dist_sel)
    list_count.append(count)

    # Upper values:
    sel_sup = dist_sup<(d)
    dist_sel_sup=dist_sup[sel_sup]
    count_sup = len(dist_sel_sup)
    list_count_sup.append(count_sup)

    # Inferior values:
    sel_inf = dist_inf<(d)
    dist_sel_inf=dist_inf[sel_inf]
    count_inf= len(dist_sel_inf)
    list_count_inf.append(count_inf)

    if dist_type=='dL':
      z_lim_dist = z_at_value(paramet.luminosity_distance, d*u.Mpc)
    if dist_type=='dZ':
      z_lim_dist = (d*H0.value)/c
    if dist_type=='dG':
      z_lim_dist = find_z(d)

    volume = (4/3)*np.pi*((d)**3)
    gama = count/volume
    gama_sup = count_sup/volume
    gama_inf = count_inf/volume

    list_gama.append(gama)
    list_gama_sup.append(gama_sup)
    list_gama_inf.append(gama_inf)
    list_dist.append(d)
    list_z.append(z_lim_dist)
    dmin = d

  return np.array(list_gama), np.array(list_gama_sup), np.array(list_gama_inf), np.array(list_dist), np.array(list_z)

# 3) Linear regression with error in both axis
def func(p,x):
    a,b=p
    return a*x+b

# 4) Axis label format
def label_form(x, pos):
    if x<1:
        xla=float(x)
        x=xla
    if x>=1:
        xla=int(x)
        x=xla
    return str(x)

# 5) Function to find slopes of gama vs dist (linear regression)
def slopes(dist_type,dist,gama,gama_lim_sup):

  # Cut related to z=1
  listateste=[]
  if dist_type == 'dL':
    for i in listaz1:
        if i>1:
            listateste.append(i)
    cut_dist = len(listaz1) - len(listateste)
  if dist_type == 'dZ':
    for i in listazz1:
      if i>1:
          listateste.append(i)
    cut_dist = len(listazz1) - len(listateste)
  if dist_type == 'dG':
    for i in listazg2:
        if i>1:
            listateste.append(i)
    cut_dist = len(listazg2) - len(listateste)

  # Linear regression for z > 1
  data_x = np.log10(dist[cut_dist:-1:])
  data_y = np.log10(gama[cut_dist:-1:])
  data_y_err = np.log10(gama_lim_sup[cut_dist:-1:]) - np.log10(gama[cut_dist:-1:])
  # Linear regression with error in both axis:
  def func(p,x):
      a,b=p
      return a*x+b
  from scipy import odr
  linear_model = odr.Model(func)
  data = odr.RealData(data_x,data_y,sy=data_y_err)
  odr = odr.ODR(data,linear_model,beta0=[1.,1.])
  out = odr.run()
  slope_g = out.beta[0]
  err_slope_g = out.sd_beta[0]
  interp_g = out.beta[1]
  err_interp_g = out.sd_beta[1]
  # Linear regression for z < 1
  from scipy import odr
  data_x = np.log10(dist[:cut_dist:])
  data_y = np.log10(gama[:cut_dist:])
  data_y_err = np.log10(gama[:cut_dist:]) - np.log10(gama_lim_sup[:cut_dist:])
  # Linear regression with error in both axis:
  def func(p,x):
      a,b=p
      return a*x+b
  linear_model = odr.Model(func)
  data = odr.RealData(data_x,data_y,sy=data_y_err)
  odr = odr.ODR(data,linear_model,beta0=[1.,1.])
  out = odr.run()
  slope_l = out.beta[0]
  err_slope_l = out.sd_beta[0]
  interp_l = out.beta[1]
  err_interp_l = out.sd_beta[1]
  return slope_g, err_slope_g, interp_g,err_interp_g, slope_l, err_slope_l, interp_l,err_interp_l


# 6) Function to create gama plots
def gama_plot(dist,gama,z,type_dist,gama_lim_sup,gama_lim_inf,slope_l,err_slope_l,interp_l,slope_g,err_slope_g,interp_g):
  plt.figure(figsize=(15,15))
  # Graph of gama_di vs d_i where i=l,z,g with bin of 200mpc and z plotted above
  fig = plt.figure()
  # Create first subplot:
  ax1 = fig.add_subplot(111)

  if type_dist=='L':
    # For the first plot:
    ax1.plot(np.log10(dist),np.log10(gama),'kx',label='$\gamma^{*}$')
    ax1.plot(np.log10(dist),interp_g_dl + slope_g_dl*np.log10(dist),'k--',label='slope(z>1) ='+str("%.2f" % round(slope_g_dl,2))+'$\pm$'+str(round(err_slope_g_dl,2)))
    ax1.plot(np.log10(dist),interp_l_dl + slope_l_dl*np.log10(dist),'k:',label='slope(z<1) ='+str("%.2f" % round(slope_l_dl,2))+'$\pm$'+str(round(err_slope_l_dl,2)))
    # Hatched area with error only on y axis
    ax1.fill_between(np.log10(dist),np.log10(gama_lim_inf),np.log10(gama_lim_sup),facecolor='grey')
    ax1.set_xlabel('$\log[d_{L}$(Mpc)]', fontsize=18)
    ax1.set_ylabel('$\log[\gamma_{L}^{*}$(Mpc$^{-3}$)]', fontsize=18)

  if type_dist=='Z':
    ax1.plot(np.log10(dist),np.log10(gama),'kx',label='$\gamma^{*}$')
    ax1.plot(np.log10(dist),interp_g_dz + slope_g_dz*np.log10(dist),'k--',label='slope(z>1) ='+str("%.2f" % round(slope_g_dz,2))+'$\pm$'+str(round(err_slope_g_dz,2)))
    ax1.plot(np.log10(dist),interp_l_dz + slope_l_dz*np.log10(dist),'k:',label='slope(z<1) ='+str("%.2f" % round(slope_l_dz,2))+'$\pm$'+str(round(err_slope_l_dz,2)))
    ax1.fill_between(np.log10(dist),np.log10(gama_lim_inf),np.log10(gama_lim_sup),facecolor='grey')
    ax1.set_xlabel('$\log[d_{Z}$(Mpc)]', fontsize=18)
    ax1.set_ylabel('$\log[\gamma_{Z}^{*}$(Mpc$^{-3}$)]', fontsize=18)

  if type_dist=='G':
    ax1.plot(np.log10(dist),np.log10(gama),'kx',label='$\gamma^{*}$')
    ax1.plot(np.log10(dist),interp_g_dg + slope_g_dg*np.log10(dist),'k--',label='slope(z>1) ='+str("%.2f" % round(slope_g_dg,2))+'$\pm$'+str(round(err_slope_g_dg,2)))
    ax1.plot(np.log10(dist),interp_l_dg + slope_l_dg*np.log10(dist),'k:',label='slope(z<1) ='+str("%.2f" % round(slope_l_dg,2))+'$\pm$'+str(round(err_slope_l_dg,2)))
    ax1.fill_between(np.log10(dist),np.log10(gama_lim_inf),np.log10(gama_lim_sup),facecolor='grey')
    ax1.set_xlabel('$\log[d_{G}$(Mpc)]', fontsize=18)
    ax1.set_ylabel('$\log[\gamma_{G}^{*}$(Mpc$^{-3}$)]', fontsize=18)

  ax1.legend(loc=3, prop={'size': 11.5})
  plt.xticks(fontsize=14)
  plt.yticks(fontsize=14)
  # Create second subplot keeping the y axis of first plot
  ax2 = ax1.twiny()
  # For the second plot
  ax2.plot(z,np.log10(gama),alpha=0)
  ax2.semilogx()
  ax2.set_xticks([0.2,0.5,1,2,3,4])
  ax2.xaxis.set_major_formatter(ticker.FuncFormatter(label_form)) # Formatting number values of second axis
  ax2.set_xlabel('z',fontsize=18)
  plt.xticks(fontsize=13)
  # Vertical line
  plt.axvline(x=1, color='k', linestyle='-')
  plt.show()
  return

# Applying function 2
listagamadl1, listagamadl1_sup, listagamadl1_inf, listadl1, listaz1 = gama_dist(dbin=200, dist=dl, dist_sup=dl_sup, dist_inf=dl_inf, dist_type='dL')
listagamadz1, listagamadz1_sup, listagamadz1_inf, listadz1, listazz1 = gama_dist(dbin=200, dist=dz, dist_sup=dz_sup, dist_inf=dz_inf, dist_type='dZ')
listagamadg2, listagamadg2_sup, listagamadg2_inf, listadg2, listazg2 = gama_dist(dbin=200, dist=dg, dist_sup=dg_sup, dist_inf=dg_inf, dist_type='dG')

# Applying function 5
slope_g_dl, err_slope_g_dl, interp_g_dl, err_interp_g_dl, slope_l_dl, err_slope_l_dl, interp_l_dl, err_interp_l_dl = slopes('dL',listadl1,listagamadl1,listagamadl1_sup)
slope_g_dz, err_slope_g_dz, interp_g_dz, err_interp_g_dz, slope_l_dz, err_slope_l_dz, interp_l_dz, err_interp_l_dz = slopes('dZ',listadz1,listagamadz1,listagamadz1_sup)
slope_g_dg, err_slope_g_dg, interp_g_dg, err_interp_g_dg, slope_l_dg, err_slope_l_dg, interp_l_dg, err_interp_l_dg = slopes('dG',listadg2,listagamadg2,listagamadg2_sup)

# Applying function 6 to create the graphs
gama_plot(listadl1,listagamadl1,listaz1,'L',listagamadl1_sup,listagamadl1_inf,slope_l_dl,err_slope_l_dl,interp_l_dl,slope_g_dl,err_slope_g_dl,interp_g_dl)
gama_plot(listadz1,listagamadz1,listazz1,'Z',listagamadz1_sup,listagamadz1_inf,slope_l_dz,err_slope_l_dz,interp_l_dz,slope_g_dz,err_slope_g_dz,interp_g_dz)
gama_plot(listadg2,listagamadg2,listazg2,'G',listagamadg2_sup,listagamadg2_inf,slope_l_dg,err_slope_l_dg,interp_l_dg,slope_g_dg,err_slope_g_dg,interp_g_dg)




           \end{lstlisting}

\chapter{Heuristic derivation of the fractal number-distance relation}

\renewcommand{\theequation}{\thechapter.\arabic{equation}}
Suppose a hierarchy as in Fig. \ref{fig:fournier}, which is a self-similar structure, discrete at order $n$. Let us suppose now that at each radius there is a certain number of objects: 
\begin{equation}
    \text{radius $d_0$ $\rightarrow$ $N_0$ objects.} \\
\end{equation}
\begin{equation}
    \text{radius $d_1$ $\rightarrow$ $N_1$ objects.}   
\end{equation}
Proportionally,
\begin{equation}
    d_1 = k d_0 \, , 
\end{equation}
\begin{equation}
    N_1 = k_{*} N_0 \, ,
\end{equation}

\noindent where $k$ and $k_{*}$ are proportionality constants. So, in general
\begin{equation}
    d_n = k^n d_0 \, ,
\lb{ap21}
\end{equation}
\begin{equation}
    N_n = k_{*}^{n} N_0 \, ,
\lb{ap22}
\end{equation}

\noindent and by writing Eq. (\ref{ap21}) we get:
\begin{equation}
    \log(d_n) = \log(k^n d_0) \, ,
\end{equation}
\begin{equation}
\log \left( \frac{d_n}{d_0} \right) = n \log(k) \, ,
\lb{ap23}
\end{equation}

\noindent as well as doing the same in Eq. (\ref{ap22}) yields,
\begin{equation}
\log \left( \frac{N_n}{N_0} \right) = n \log(k_*).
\lb{ap24}
\end{equation}

\noindent Dividing Eq. (\ref{ap24}) by Eq. (\ref{ap23}) we have:
\begin{equation}
    \frac{\log(N_n/N_0)}{\log(d_n/d_0)} = \frac{n \log(k_*)}{n \log(k)} \, ,
\end{equation}
\begin{equation}
    \log \left( \frac{N_n}{N_0} \right) = \frac{\log(k_*)}{\log(k)} \log \left( \frac{d_n}{d_0} \right) \, ,
\end{equation}

\noindent and then we have that, where 
\begin{equation}
    D = \frac{\log(k_*)}{\log(k)} \, ,
\end{equation}
\begin{equation}
    \log \left( \frac{N_n}{N_0} \right) = \log \left( \frac{d_n}{d_0} \right) ^D \, .
\end{equation}

\noindent Removing the logarithms we get:
\begin{equation}
    \frac{N_n}{N_0} = \left( \frac{d_n}{d_0} \right)^D \, ,
\end{equation}
\begin{equation}
    N_n = \frac{N_0 d_n^D}{d_0^D} \, ,
\end{equation}
\noindent and defining $B$ as:
\begin{equation}
    B = \frac{N_0}{d_0^D} \, ,
\end{equation}
\noindent we subsequently attain
\begin{equation}
    N_n = B d_n^D \, .
\end{equation}
\noindent Thus, in general, for a smooth proportionality we reach to the number distance relation for a self-similar fractal structure.
\begin{equation}
    N = B d^D \, .
\end{equation}

\pagebreak
\phantomsection
\addcontentsline{toc}{chapter}{Bibliography}
\bibliography{references}
\renewcommand{\thechapter}{\arabic{chapter}}
\end{document}